\documentclass{article}

\usepackage{arxiv}

\usepackage[utf8]{inputenc} 
\usepackage[T1]{fontenc}    
\usepackage{url}            
\usepackage{booktabs}       
\usepackage{amsfonts}       
\usepackage{nicefrac}       
\usepackage{microtype}      
\usepackage{lipsum}		    
\usepackage{graphicx}
\usepackage{doi}

\usepackage{orcidlink}
\hypersetup{hidelinks}
\usepackage{mathrsfs} 
\usepackage{amssymb, amsthm, amsmath}
\usepackage{ulem}

\newtheorem{definition}{Definition}

\title{Laterally constrained low-rank seismic data completion via cyclic-shear transform}

\author{ 
    \orcidlink{0000-0002-6501-752X} David Vargas
    \thanks{Formerly, Department of Earth Sciences, Utrecht University, The Netherlands} \\
	Shearwater GeoServices \\
	Gatwick, United Kingdom \\ 
	\texttt{dvargas@shearwatergeo.com} \\
	\And
    \orcidlink{0000-0001-7405-1490} Ivan Vasconcelos
    \footnotemark[1] \\
	Shearwater GeoServices \\
	Gatwick, United Kingdom \\ 
	\texttt{ivasconcelos@shearwatergeo.com} \\
	\And
    Nick Luiken \\
	KAUST \\
	Thuwal, Kingdom of Saudi Arabia \\
	\texttt{nicolaas.luiken@kaust.edu.sa} \\
	\And
    \orcidlink{0000-0003-0020-2721} Matteo Ravasi \\
	KAUST \\
	Thuwal, Kingdom of Saudi Arabia \\
	\texttt{matteo.ravasi@kaust.edu.sa} \\
}

\renewcommand{\shorttitle}{Laterally constrained LRMC via cyclic-shear transform}

\hypersetup{
pdftitle={Laterally constrained low-rank seismic data completion via cyclic-shear transform},
pdfsubject={A pre-print for the arxiv style},
pdfauthor={David Vargas, Ivan Vasconcelos, Nick Luiken and Matteo Ravasi},
pdfkeywords={low-rank matrix completion, seismic interpolation},
}

\begin{document}
\maketitle

\begin{abstract}
    A crucial step in seismic data processing consists in reconstructing the wavefields at spatial locations where faulty or absent sources and/or receivers result in missing data. Several developments in seismic acquisition and interpolation strive to restore signals fragmented by sampling limitations; still, seismic data frequently remain poorly sampled in the source, receiver, or both coordinates. An intrinsic limitation of real-life dense acquisition systems, which are often exceedingly expensive, is that they remain unable to circumvent various physical and environmental obstacles, ultimately hindering a proper recording scheme. In many situations, when the preferred reconstruction method fails to render the actual continuous signals, subsequent imaging studies are negatively affected by sampling artefacts. A recent alternative builds on low-rank completion techniques to deliver superior restoration results on seismic data, paving the way for data kernel compression that can potentially unlock multiple modern processing methods so far prohibited in 3D field scenarios. In this work, we propose a novel transform domain revealing the low-rank character of seismic data that prevents the inherent matrix enlargement introduced when the data are sorted in the midpoint-offset domain and develop a robust extension of the current matrix completion framework to account for lateral physical constraints that ensure a degree of proximity similarity among neighbouring points. Our strategy successfully interpolates missing sources and receivers simultaneously in synthetic and field data. 
\end{abstract}


\section{Introduction}\label{sec_1}
Modern seismic technology aims at building accurate subsurface images of stratigraphic and structural parameters describing different geological properties of the specific region of interest. Reconstructing 3D earth models relies on several processing algorithms that commonly expect ideal, fully sampled seismic responses along both the source and the receiver planes. Examples of such processing methods include multiple prediction and elimination \cite{Verschuur1992, Dragoset2010}, interferometric redatuming \cite{Schuster2006, Wapenaar2011, Ravasi2015}, reverse-time migration \cite{Baysal1983}, and waveform inversion \cite{Sirgue2004}, to name a few. To accurately restore the seismic response, digital processing relies on a discretised acquisition scheme that obeys the Nyquist-Shannon sampling theorem \cite{Nyquist1928}. According to the sampling theorem, the sampling rate must comply with the specific phase velocity of the propagating wavefield for a given source frequency and should at least be twice the maximum frequency of the physical signal. In general, seismic data are well sampled in time; however, it is difficult to properly fulfil the prescribed source/receiver carpet equidistant spacing required to reconstruct the underlying alias-free signal. Some of the aspects driving such limitations include multiple physical, economic, or environmental constraints in which seismic exploration is conducted. In view of the adverse sampling scenario, various interpolation techniques have experienced an increasingly important role in seismic processing under the promise of delivering a densely sampled dataset that ultimately mitigates the intrinsic restrictions of limited spatial observation nodes.

A suboptimal regular-sampling strategy results in data aliasing, whereas a signal registered at random locations manifests itself with a noise-polluted spectrum in a certain domain where the signal has a sparse representation \cite{Cao2011}. From this perspective, one distinguishes two distinct signal reconstruction categories based on the nature of the problem and how sampling is conducted. The first group rely on nonaliased low frequencies to build prediction filters in the frequency-space (f-x) or the frequency-wavenumber (f-k) domain. Such strategies use the slowest linear events to interpolate aliased data at high frequencies \cite{Spitz1991, Necati2003, Xu2005}. The second class corresponds to compressed sensing (CS) techniques leveraging the assumption that data are well represented on certain sparsity-promoting domains, therefore, allowing a seamless $L1$ norm regularized inversion \cite{Donoho2006}. Typical reconstruction domains are introduced through the Radon \cite{Kabir1995, Ibrahim2018}, Fourier \cite{Sacchi1998, Naghizadeh2011}, Curvelet \cite{Herrmann2008, Naghizadeh2010}, Seislet \cite{Fomel2010}, Shearlet \cite{Guo2007}, or Dreamlet \cite{Wu2013, Benfeng2015} transforms. Alternatively, propagation principles via wave continuation are used for interpolation \cite{Ronen1987, Stolt2002} along with methods based on wave-front attributes that aim at including structural information of the subsurface \cite{Yujiang2017}. An active field of research concerns AI-based methods, in particular, machine learning algorithms where multiple architectures have been recently proposed for seismic data reconstruction \cite{Wang2019, Liu2019, Brandolin2022}. 

Recent advances consider wavefield interpolation via reduced-rank approximations, building on the assumption that data are structured and can be encapsulated in a low-dimensional space via rank-revealing transformations. Notably, algorithms based on multichannel singular spectrum analysis (MSSA) \cite{Bahia2019, Carozzi2021, Rongzhi2022} use projection-onto-convex-sets \cite{Abma2006} to complete a large Hankel matrix encoding the seismic signals. This family of methods expand the ideas of CS to structured matrices, where the fast decay of singular values serves as an analogue of sparsity, i.e., the data are fully explained with only a few non-zero singular values. A prominent CS technique of particular importance is Low-Rank matrix completion (LRMC), typically used for seismic interpolation in the frequency-space domain, where data are organized to reveal its low-rank nature. Some examples conducted on seismic data include tensor completion methods \cite{Ely2015}, higher-order singular value decomposition \cite{Kreimer2013}, SVD-free low-rank matrix factorization \cite{Aravkin2014}, rank minimization via alternating optimization \cite{Kumar2015}, and low-rank matrix completion with texture-patch mapping \cite{Ma2013}. 

In the source-receiver $(\omega, x_{r}, x_{s})$ domain, seismic data exhibit a high-rank structure by virtue of the significant energy contribution of zero-offset data allocated along the main diagonal entries that progressively decays towards the off-diagonal directions as a result of geometrical spreading. Interestingly, the same data are low-rank in the midpoint-offset domain $(\omega, m, h)$ since the mapping rule is such that the original array undergoes a $45$ deg rotation. Typically, seismic data are completed in the midpoint-offset domain, where the presence of noise and missing sources/receivers increase the matrix rank; hence, an iterative shrinkage of the singular values results in matrix restoration. Although singular value-free MC is suitable for large-scale problems, such methods require prior knowledge of the actual rank \cite{Kumar2015}, and there is no clear answer as to how this parameter varies with frequency and data complexity. Consequently, some instances require explicit SVD computations. Additionally, a $45$ deg rotation enlarges the matrix with zero entries, duplicating the array size for the particular $n \times n$ case (square matrix) $n^{2} \rightarrow 2 n^{2}$. In general, a $m \times n$ source-receiver matrix turns into a $(m+n)^{2}/2$ midpoint-offset counterpart.

For large-scale problems, the memory burden may hinder MC. Our contribution focuses on mitigating the impact of enlarging a matrix in the midpoint-offset domain. We introduce a novel, mathematically inspired, transform promoting low rank in a cyclic-shear domain that avoids data sorting in the midpoint-offset domain. This alternative transformation is implemented through circular permutations of the data and proves to be effective for simultaneous source-receiver interpolation. In this letter, we approach the simultaneous reconstruction of missing sources and receivers and regard interpolation along a single side, either source or receiver, as a particular case. We also show that additional regularization enforcing explicit proximity similarity among neighbouring sources and receiver benefit the inversion and guarantees a stable solution for high decimation rates. The effectiveness of our matrix completion approach is evaluated with an extensive set of numerical tests in synthetic and field data.

\section{Low-rank matrix estimation}\label{sec_2}
This section presents the underlying theoretical principles driving low-rank MC for seismic wavefield interpolation. Completing a matrix $\mathbf{X}$ is a process that naturally introduces compressive sensing ideas, on the one hand, while requiring the data structure to physically adhere to a degree of local similarity across neighbouring measurements, on the other. The first aspect of this notion deals with the mathematical principles of CS, meaning that, in principle, a sparse signal can be recovered from fewer samples than the Nyquist sampling theorem prescribes, as long as the signal is incoherently sampled \cite{Donoho2006, Candes2008}. Therefore, in this work, we rely on natural signals with concise domain representations that exhibit as few nonzero entries as possible while ensuring the sampling scheme is a random process that guarantees a low correlation between any two elements of the restriction operator. The form of sparsity considered in the MC problem is related to the low-rank structure of the desired solution; in other words, the prior belief that the singular values distribution of $\mathbf{X}$ is a sparse signal. The second aspect encodes the physics of wave propagation throughout continuous media that results in smooth variations across neighbouring observation points, except for areas of sharp discontinuities. 

\subsection{Notation}\label{sec_2.1}
Throughout the paper, we reserve lowercase boldface letters $\mathbf{x}$ to denote vectors and uppercase boldface letters $\mathbf{X}$ matrices. Calligraphic uppercase fonts $\mathcal{R}$ indicate operators. Similarly, matrix elements are denoted $X_{ij}$ while vector entries $x_{i}$. Any scalar is represented by lowercase Greek letters $\lambda$. The singular value decomposition (SVD) of an $m \times n$ matrix $\mathbf{X} = \mathbf{U} \mathbf{\Sigma} \mathbf{V}^{\ast}$ of rank $r \le \text{min}\{m, n\}$ is given in terms of matrix $\mathbf{U} = [\mathbf{u}_{1}, \cdots, \mathbf{u}_{r}] \in \mathbb{R}^{m \times r}$ of right singular vectors $\mathbf{u}_{i}$ (orthonormal columns), a diagonal matrix $\mathbf{\Sigma} = \mbox{diag}(\sigma_{i}(\mathbf{X})) \in \mathbb{R}^{r \times r}$ with singular values $\sigma_{i}$ organized in descending order, and matrix $\mathbf{V} = [\mathbf{v}_{1}, \cdots, \mathbf{v}_{r}] \in \mathbb{R}^{n \times r}$ of left singular vectors $\mathbf{v}_{i}$. Likewise, $\left\Vert \mathbf{X} \right\Vert_{F} = \sqrt{\sum_{i=1}^{m} \sum_{j=1}^{n} \left| X_{ij} \right|^{2}}$ is the Frobenius norm of the same matrix, and $\| \mathbf{X} \|_{*} = \| \mathbf{\sigma}(\mathbf{X}) \|_{1} = \sum_{i=1}^n |\sigma_{i}|$ its nuclear norm, sometimes also known as trace norm. 

\subsection{Cyclic shift}\label{sec_2.2}
We regard $S_{n}$ as the permutation group defined on the finite set $A_{n} = \{a_{1}, a_{2}, ..., a_{n}\}$ for any positive integer $n \geq 1$ which contains all arrangements of degree $n$. In particular, we introduce the transition of a specific element $a_{i+1} \in A_{n}$ to the adjacent location $\mathbf{\pi}: a_{i+1} \mapsto a_{i}$ and call the map $\pi$ cyclic if $\mathbf{\pi}: a_{1} \mapsto a_{n}$. The following definition generalizes the cyclic map so that the shift accepts a step of length $k$.  
\vspace{4pt}
\begin{definition}[$k$-cyclic permutation]
Let the finite set $A_{n} = \{a_{1}, a_{2}, ..., a_{n}\}$ of $n \geq 1$ elements with entry indices the positive integer set $\mathbb{Z}^{+}_{n} = \{1,2,...,n\}$ be an initial state of the permutation group $S_{n}$. The circular or cyclic permutation of length $k$ is the bijective function $\mathbf{\pi}: S_{n} \to S_{n}$ defined by the index map
\begin{align*}
    a_{\mathbf{\pi}_{k}(i)}        \: &= \: a_{(i - k) \; \text{mod} \; n} \: , \quad i \in Z^{+}_{n} , \\
    a_{\mathbf{\pi}^{\ast}_{k}(i)} \: &= \: a_{(i + k) \; \text{mod} \; n} \: , \quad i \in Z^{+}_{n} .
\end{align*}
\label{def_1}
\end{definition}

In forward mode, the cyclic shift $\mathbf{\pi}$ rearranges the set $A_{n}$ by shifting to the right each entry $k$-times to the next position while allocating the last entry to the $k$th position. Using Cauchy's two-line notation for $k=1$, it is:
\begin{equation*}
    \mathbf{\pi}_{1} = 
    \begin{pmatrix}
        a_{1} & a_{2} & ... & a_{n-2} & a_{n-1} & a_{n} \\
        a_{n} & a_{1} & a_{2} & ... & a_{n-2} & a_{n-1} 
        \end{pmatrix} \; .
\end{equation*}
On the other hand, in adjoint mode, this operator moves the first entry to the last position, and shifts to the left each entry by a single position, i.e., 
\begin{equation*}
    \mathbf{\pi}^{\ast}_{1} = 
    \begin{pmatrix}
        a_{1} & a_{2} & a_{3} & ... & a_{n-1} & a_{n} \\
        a_{2} & a_{3} & ... & a_{n-1} & a_{n} & a_{1}
        \end{pmatrix} \; .
\end{equation*}
There are no restrictions in $l$ multiple applications of the map $\pi$ on a set; therefore, it can repeatedly act on $A_{n}$, resulting in an $kl$-cyclic shift of the tuple. Note that the circular shift excludes the location exchange of adjacent elements.
    
\begin{figure}[!t]
    \centering
    \includegraphics[scale=0.09]{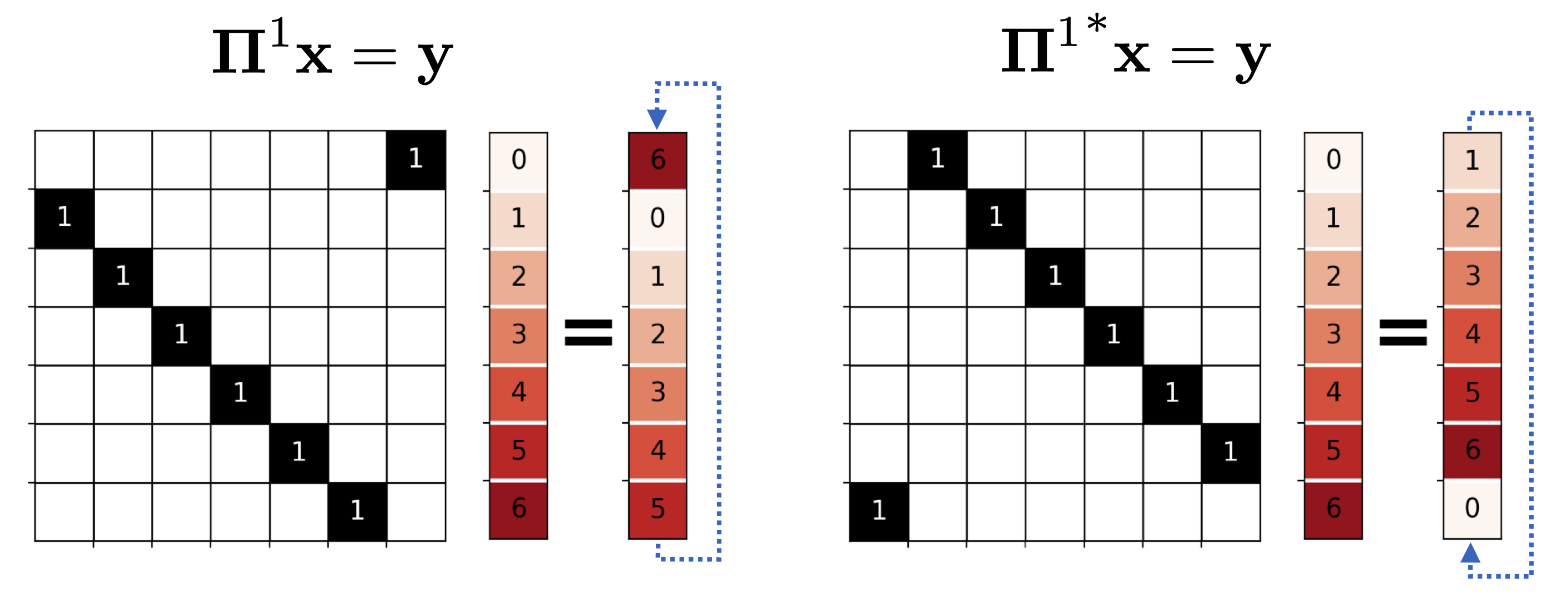}
    \caption{
    Illustration of the $7 \times 7$ cyclic shift matrix $\mathbf{\Pi}$ acting on a generic $7 \times 1$ vector $\mathbf{x}$ according to the rule $[\mathbf{\Pi}^{1} \mathbf{x}]_{i} = x_{\pi_{1}}$ in forward mode, and $[{\mathbf{\Pi}^{1}}^{\ast} \mathbf{x}]_{i} = x_{\pi^{\ast}_{1}}$ in adjoint mode. This binary matrix cyclically permutes the vector elements one position downwards while its adjoint does so in the opposite direction.
    }
    \label{fig_1}
\end{figure}

Since we consider operators $\mathbf{\Pi} \in \mathbb{R}^{n \times n}$ acting on vectors $\mathbf{x} \in \mathbb{R}^{n}$, it is convenient to write the cyclic shift in matrix form.

\vspace{4pt}
\begin{definition}[Cyclic shift matrix of degree $k$]
For $i, j \in \mathbb{Z}^{+}_{n}$, let $\{\mathbf{e}_{i} \in \mathbb{R}^{n}\}$ be the canonical basis in the $n$-dimensional Euclidean space, and let the identity $\mathbf{I}_{n} = [\mathbf{e}_{1}, \mathbf{e}_{2}, ... , \mathbf{e}_{n-1}, \mathbf{e}_{n}]$ be the zero-order cyclic shift matrix $\mathbf{\Pi}^{0} \in \mathbb{R}^{n \times n}$. The cyclic shift matrix $\mathbf{\Pi}^{k} \in \mathbb{R}^{n \times n}$ of degree $k$ and its adjoint are given by the recursions
\begin{align*}
     \mathbf{\Pi}^{k}_{i, j}         \: &= \: \mathbf{\Pi}^{k-1}_{\mathbf{\pi}_{1}(i), j}        \: , \\
    {\mathbf{\Pi}^{k}_{i, j}}^{\ast} \: &= \: \mathbf{\Pi}^{k-1}_{\mathbf{\pi}^{\ast}_{1}(i), j} \: , 
\end{align*}
where $k \in \mathbb{Z}^{+}_{n}$ and $\pi_{1}$ is the $1$-cyclic permutation as in the \textit{Definition} \ref{def_1}.
\label{def_2}
\end{definition}

Owing that both column and row spaces of the cyclic shift are orthonormal bases; hence, the matrix $\mathbf{\Pi}^{k}$ is orthogonal ${\mathbf{\Pi}^{k}}^{\ast} \mathbf{\Pi}^{k} = \mathbf{\Pi}^{k} {\mathbf{\Pi}^{k}}^{\ast} = \mathbf{I}_{n}$, or equivalently, the inverse equals the algebraic adjoint, i.e.,  ${\mathbf{\Pi}^{k}}^{-1} = {\mathbf{\Pi}^{k}}^{\ast}$.

In particular, moving each column in the identity matrix by one position to the left using the cyclic permutation yields the first-order cyclic shift matrix,
\begin{align*}
\mathbf{\Pi}^{1} = \begin{bmatrix}
        0      & 0 & 0      & \ldots & 1       \\
        1      & 0 & 0      &        & 0       \\
        0      & 1 & 0      & \ddots & 0       \\
        \vdots &   & \ddots & \ddots & \vdots  \\
        0      & 0 & \ldots & 1      & 0
        \end{bmatrix}, \: 
{\mathbf{\Pi}^{1}}^{\ast} = \begin{bmatrix}
        0      & 1 & 0      & \ldots & 0       \\
        0      & 0 & 1      &        & 0       \\
        0      & 0 & 0      & \ddots & \vdots  \\
        \vdots &   & \ddots & \ddots & 1       \\
        1      & 0 & 0      & \ldots & 0
        \end{bmatrix} .
\end{align*}
Once the zero-order matrix $\mathbf{\Pi}^{0} = \mathbf{I}_{n}$ is defined, higher-order circular shifts follow from the \textit{Definition} \ref{def_2}. The action of $\mathbf{\Pi}^{1}$ and its adjoint ${\mathbf{\Pi}^{1}}^{\ast}$ on the vector $\mathbf{x} \in \mathbb{R}^{n}$ is such that 
 \begin{align*}
    {\mathbf{\Pi}^{1}}^{}     [x_{1}, x_{2}, ..., x_{n-1}, x_{n}]^{T} \: &= \: [x_{n}, x_{1}, ...,x_{n-2}, x_{n-1}]^{T} , \\ 
    {\mathbf{\Pi}^{1}}^{\ast} [x_{1}, x_{2}, ..., x_{n-1}, x_{n}]^{T} \: &= \: [x_{2}, x_{3}, ...,x_{n}, x_{1}]^{T} .
\end{align*}
Figure \ref{fig_1} illustrates the action of the forward and adjoint operation of a $7 \times 7$ circular shift matrix on an arbitrary $7 \times 1$ vector. The circular transformation can also use fractional shifts of arbitrary precision by introducing a fractional order power of $\mathbf{\Pi}$ or directly via phase shifts in the space-wavenumber domain. We explore this property in a companion paper.  

\subsection{Low-Rank models for Matrix completion}\label{sec_2.3}
We are interested in restoring a matrix from a fragmented set of entries observations at given random positions. In such case, the standard protocol considers the measurement operator defined by the linear map $\mathcal{R}: \mathbb{R}^{m \times n} \to \mathbb{R}^{h}$, with matrix representation $\mathcal{R}(\mathbf{X}) = \mathbf{R} \mbox{vec}(\mathbf{X})$, that selects a small yet representative number of samples $h < mn$ and connects the fully-sampled signal $\mathbf{x} = \mbox{vec}(\mathbf{X})$ with available partial observations $\mathbf{y}$. Then, the undersampled data are expressed as the missing entry reconstruction problem
\begin{equation}
    \mathbf{y} = \mathbf{R} \mathbf{x} + \mathbf{n} \:,
    \label{eq1}\tag{1}
\end{equation}
where $\mathbf{x} = [\mathbf{x}^{T}_1, \mathbf{x}^{T}_2, ..., \mathbf{x}^{T}_n]^{T} \in \mathbb{R}^{mn}$ is the one-dimensional vectorized version of the matrix of interest resulting from stacking its columns on top of one another, i.e., $\mathbf{x} = \mbox{vec}(\mathbf{X})$, $\mathbf{y} = [y_1, y_2, ..., y_h]^{T} \in \mathbb{R}^{h}$ is the observed fragmented signal with missing entries, and $\mathbf{n} = [n_1, n_2, ..., n_h]^{T} \in \mathbb{R}^{h}$ is a noise vector. Note that $\mathbf{R} \in \mathbb{R}^{h \times mn}$ is a sensing matrix extracting a subset of elements from the input vector at specific locations, formally defined in terms of the Kronecker delta \cite{BinLiu2004}
\begin{equation}
    R_{i j} = \delta_{h(i) j} = 
    \begin{cases} 
        1, & \mbox{if } h(i)=j \\ 0, & \mbox{if } h(i) \neq j 
    \end{cases} \:,
    \label{eq2}\tag{2}
\end{equation}
with $ \mathcal{H} = \{h(i) \mid i = 1, 2, ..., H\}$, the set of indices indicating the entries of the samples to retain. Recovering the matrix $\mathbf{X}$ from $h < mn$ samples is an ill-conditioned problem, challenging to solve without introducing additional prior information on the nature of the sought-after solution. Among all possible arrays, we restrict the solution so that the reconstruction exhibits a low-dimensional structure or, equivalently, a low-rank configuration. This assumption is based on the premise that adjacent entries share a degree of similarity, a consideration that arises in many practical instances \cite{Candes2010A, Davenport2016}. Since the rank minimization problem is intractable \cite{Natarajan1995}, an alternative option considers the constrained optimization problem:
\begin{equation}
    \min_{\mathbf{x}} \: \| \mathbf{X} \|_{*} \quad \text{s.t.} \quad \| \mathbf{R} \mathbf{x} - \mathbf{y} \|_{2}^{2} \leq \epsilon \:, 
    \label{eq3}\tag{3}
\end{equation}
where the nuclear norm serves as a convex relaxation for rank minimization \cite{Recht2010, Candes2010B, Candes2012} and the desired matrix $\mathbf{X}$ is consistent with the measurements up to the tolerance level $\epsilon$. Equivalently, moving the constraint into the objective function, MC is formulated as the regularized least-squares problem:
\begin{equation}
    \min_{\mathbf{x}} \: \| \mathbf{R} \mathbf{x} - \mathbf{y} \|_{2}^{2} \:+\: \mu \| \mathbf{X} \|_{*} \:.
    \label{eq4}\tag{4}
\end{equation}
Although the MC program \eqref{eq4} effectively finds the lowest rank estimate $\mathbf{X}$ offering superior performance, the trace norm evaluation requires multiple singular value decompositions (SVD), an operation that, for large-scale practical applications, is bounded by the SVD numerical complexity $\mathcal{O}(n^{3})$ and the implicit storage requirements. To overcome this issue, SVD-free minimization techniques emerged as an alternative that meets the expectation of finding a low-rank matrix while avoiding the scalability burden by reformulating the inversion in terms of the product $\mathbf{X} = \mathbf{U}\mathbf{V}^{\ast}$ with $\mathbf{U} \in \mathbb{R}^{m \times r}$ and $\mathbf{V} \in \mathbb{R}^{n \times r}$, such that the low-rank factorization model
\begin{equation}
    \min_{\mathbf{U, V}} \: \tfrac{1}{2} (\| \mathbf{U} \|^{2}_{F} + \| \mathbf{V} \|^{2}_{F}) \quad \text{s.t.} \quad \| \mathbf{R} \mathbf{x} - \mathbf{y} \|_{2}^{2} \leq \epsilon \:, 
    \label{eq5}\tag{5}
\end{equation}
or its damped least squares version,
\begin{equation}
    \min_{\mathbf{U, V}} \: \| \mathbf{R} \mathbf{x} - \mathbf{y} \|_{2}^{2} \:+\: \mu (\| \mathbf{U} \|^{2}_{F} + \| \mathbf{V} \|^{2}_{F})
    \label{eq6}\tag{6}
\end{equation}
is proposed as an analogue of MC \cite{Recht2013} with $\mathbf{x} = \mbox{vec}(\mathbf{U}\mathbf{V}^{\ast})$. This low-dimensional factor model is driven by the nuclear norm decomposition $\| \mathbf{X} \|_{*} \leq \tfrac{1}{2} (\| \mathbf{U} \|^{2}_{F} + \| \mathbf{V} \|^{2}_{F})$ \cite{Srebro2004, Jasson2005} and predicts an accurate solution up to $\mbox{rank}(\mathbf{X}) \leq r$. The problem is solved using alternating minimisation by iteratively fixing one factor and optimising over the other. Even though \eqref{eq6} is non-convex, each sub-problem is convex in alternating minimisation. Given a proper initialisation of $\mathbf{U}$ and $\mathbf{V}$, an optimal solution is guaranteed on the condition that $\mbox{rank}(\mathbf{X})$ is significantly smaller than $r$ \cite{Burer2003, Burer2005}. Alternatively, the max-norm can also be used as a regularizer enforcing low-rank structure \cite{Lee2010}
        
\begin{figure}[!t]
    \centering
    \includegraphics[scale=0.47]{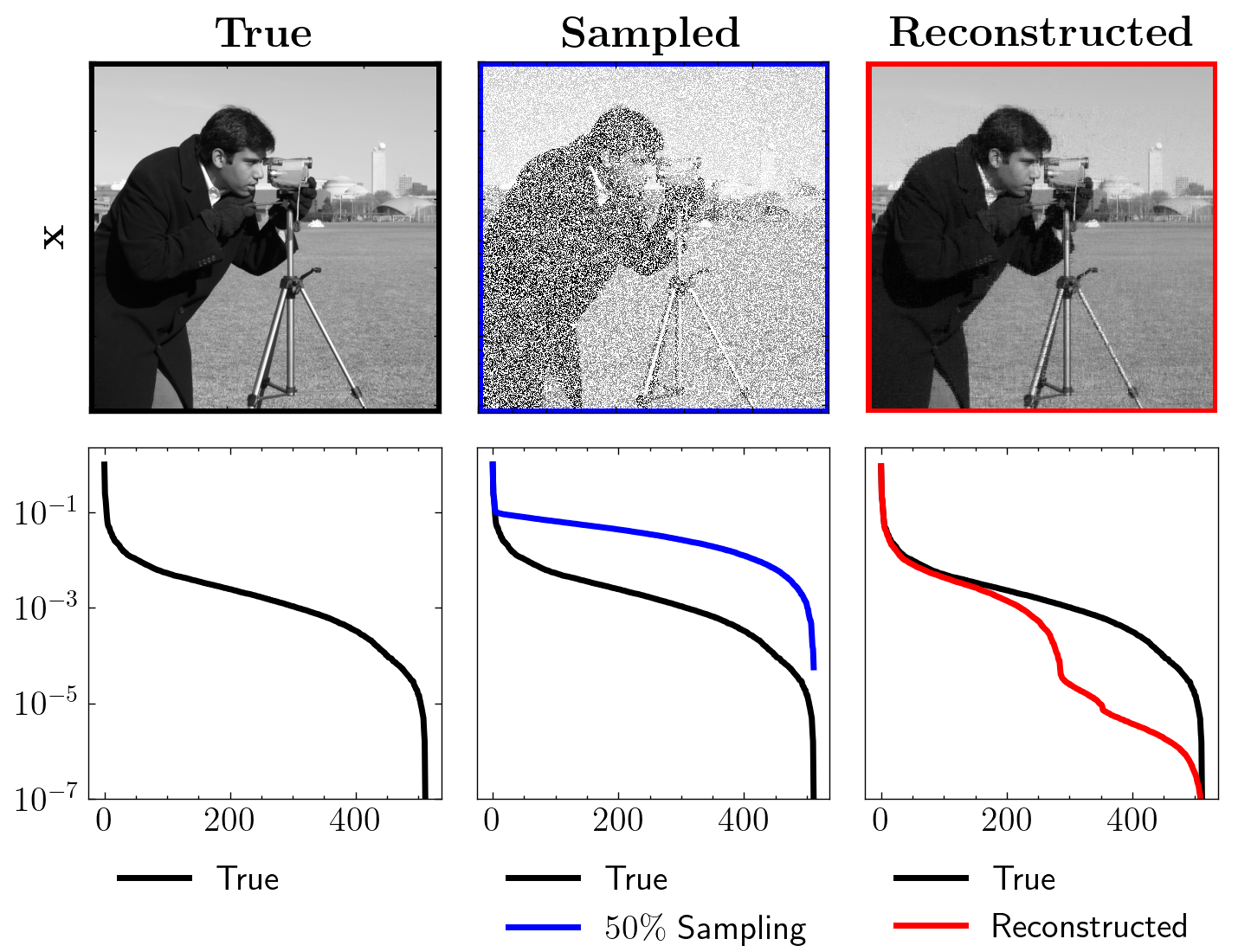}
    \caption{
    Digital image completion with low-rank techniques. Recovering a $512 \times 512$ matrix from $50\%$ of the data acquired randomly results in a $23.11 \: \text{dB}$  signal-to-noise ratio image with minimum perceptual loss. The top panels depict the original, sampled and recovered image. Conversely, the bottom panels show the evolution of the singular values matching each image as indicated by the frames' colour.
    }
    \label{fig_2}
\end{figure}

\subsection{Matrix recovery algorithms}\label{sec_2.4}
Besides constraining the restored matrix to be low-rank, having access to the actual rank in advance determines the nature of the completion algorithm. When the rank is unknown, a simple approach for the unconstrained convex optimization program \eqref{eq4} is the singular value thresholding (SVT) subroutine \cite{Cai2010}. In SVT, the proximal operator of $\mathbf{\hat X}$ associated with the nuclear norm $f(\mathbf{X}) = \| \mathbf{X} \|_{*}$
\begin{equation}
    \mbox{prox}_{\tau f}(\mathbf{\hat X}) = \mbox{arg} \min_{\mathbf{X}} \: f(\mathbf{X}) + \frac{1}{2\tau} \big\| \mathbf{X} - \mathbf{\hat X} \big\|_{F}^{2}
    \label{eq7}\tag{7}
\end{equation}
minimizes the trace norm while enforcing $\mathbf{X}$ to remain in the vicinity of $\mathbf{\hat X}$. Then, the proximal gradient recursion for the MC problem \eqref{eq4} is expressed as \cite{Parikh2014, Beck2017}:
\begin{equation}
    \mathbf{X}_{k+1} = \mathcal{D}_{\lambda_{k}}(\mathbf{X}_{k} - \tau_{k} \mathcal{R}^{\ast}(\mathcal{R} (\mathbf{X}_{k}) - \mathbf{y})) \: ,
    \label{eq8}\tag{8}
\end{equation}
with $k$ the iteration counter, $\tau_{k}$ an appropriate step size, and $\lambda_{k}$ a parameter balancing data fidelity against low-rank structure in $\mathbf{X}$. Note that the proximal operator for the nuclear norm is well-defined by the singular value thresholding map
\begin{align*}
    \mathcal{D}_{\lambda_{k}}(\mathbf{X}) &= \mathbf{U} \mathcal{T}_{\lambda_{k}}(\mathbf{\Sigma}) \mathbf{V}^{\ast}  \:, \\
    \quad \mathcal{T}_{\lambda_{k}}(\mathbf{\Sigma}) &= \mbox{diag}( \mbox{max}(\sigma_{i} - \lambda_{k}, 0)) \:.
    \label{eq9}\tag{9}
\end{align*}
Such operation effectively restricts $\mbox{rank}(\mathbf{X})$ by setting small singular values $\sigma_{i} < \lambda_{k}$ to zero, whereas large values $\sigma_{i} \geq \lambda_{k}$ are gradually shrunk towards zero, i.e., soft thresholding operation. Ultimately $\mathcal{D}_{\lambda_{k}}$ in \eqref{eq8} finds a sparse vector of singular values along with the corresponding basis in which the representation is low-rank \cite{Cai2010}.   
    
\begin{figure}[!b]
    \centering
    \includegraphics[scale=0.080]{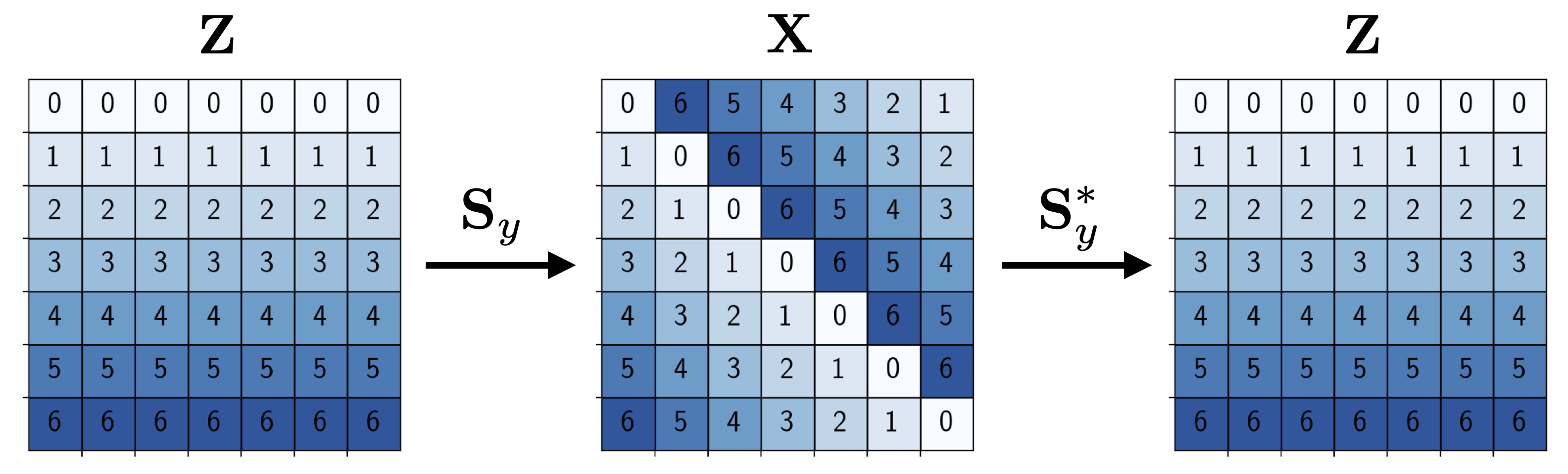}
    \caption{
    The cyclic-shear transformation $\mathbf{S}_{y}$ and its effect on a $7 \times 7$ rank-one matrix $\mathbf{Z} = \mathbf{1} \otimes \mathbf{c}_{n}$. This map operates on the input array's columns and cyclically shifts each column's entries by a certain offset according to the relation $k(j) = \lfloor \beta j \rfloor$ with $\beta = 1$. The new matrix $\mathbf{X}$ is a high-rank circulant with the first column remaining unchanged, the second rolling one position downwards, and the third and fourth shifting two and three places, respectively.
    }
    \label{fig_3}
\end{figure}

Other algorithms for MC that do not require rank information include the alternating direction method of multipliers (ADMM) \cite{Boyd2011, Glowinski2014}, the iteratively reweighted least squares method (IRLS) \cite{Chartrand2008, Fornasier2011, Karthik2012}, and the Primal-Dual Hybrid Gradient (PDHG) method \cite{Zhu2008, Esser2010, Chambolle2011, goldstein2015}, to name a few. If an estimate of the rank is known, the Projected gradient descent algorithm \cite{Recht2013}
\begin{align*}
    \mathbf{U}_{k+1} = \mathcal{P}_{\eta}(\mathbf{U}_{k} - \tau_{k} \nabla_{\eta}f(\mathbf{U}_{k}, \mathbf{V}_{k})) \: , 
    \label{eq10.1}\tag{10.1} \\
    \mathbf{V}_{k+1} = \mathcal{P}_{\xi}(\mathbf{V}_{k} - \tau_{k}  \nabla_{\xi}f(\mathbf{U}_{k}, \mathbf{V}_{k})) \:, 
    \label{eq10.2}\tag{10.2}
\end{align*}
use the proximal-point map $\mathcal{P}$ to simultaneously optimise $\mathbf{U}$ and $\mathbf{V}$. A closely related scheme fixes one of the factors while optimising over the other, resulting in the alternating minimisation step \cite{Haldar2009, Jain2012, Tanner2016}
\begin{align*}
    \mathbf{U}_{k+1} \: &= \: \mbox{arg} \min_{\mathbf{U}} \: \| \mathbf{R} \mbox{vec}(\mathbf{U}\mathbf{V}^{\ast}_{k}) - \mathbf{y} \|_{2}^{2} \: ,
    \label{eq11.1}\tag{11.1} \\
    \mathbf{V}_{k+1} \: &= \: \mbox{arg} \min_{\mathbf{V}} \: \| \mathbf{R} \mbox{vec}(\mathbf{U}_{k+1}\mathbf{V}^{\ast}) - \mathbf{y} \|_{2}^{2} \: .
    \label{eq11.2}\tag{11.2}
\end{align*}
Alternative options belonging to this class include low-rank matrix fitting (LMaFit) \cite{Wen2012}, atomic decomposition for minimum rank approximation (ADMiRA) \cite{K_Lee2010}, and truncated nuclear norm minimization TNNM, which apply a singular value hard thresholding at the projection step \cite{Jain2010, Hu2013, Tanner2013}.  
In this study, we assume the rank to be unknown and opt for the primal-dual algorithm \cite{Chambolle2011} to complete seismic data arrays. Still, some concepts may well apply to cases where an estimation of the rank is available. To illustrate the efficiency of nuclear norm minimization, we conduct a first numerical experiment dealing with a simple image reconstruction problem where missing entries are uniformly distributed at random. Ensuring the remaining entries carry comparable information from their immediate neighbouring entries is essential in matrix completion. Therefore, to prevent an unbalanced sample removal, we apply a random mask covering the original image and remove 50$\%$ of the data. The original image, its sampled version, and the recovered counterpart are presented in Figure \ref{fig_2} along with their corresponding singular value spectrum. This figure confirms the successful recovery of a high-fidelity image representation of the true data. Effectively, random missing pixels in the sampled image lead to an overall increment in singular values compared to that observed in the singular values spectrum associated with the original image. Simply put, the matrix completion program \eqref{eq8} minimises the fidelity term in the least-squares sense while balancing the low-rank character of the proposed solution through the trade-off damping parameter $\lambda_{k}$. As iterations progress, the singular values spectrum progressively shrinks until a stopping criterion is reached.  

\section{Harnessing low-rank structure}\label{sec_3}
In seismic technology, the subsurface response to a set of sources is recorded at receiver locations for a given time interval. Standard 2D-seismic surveys with data sorted in the common shot-receiver domain result in a temporal 3D array $\mathbf{X} \in \mathbb{R}^{nt \times nr \times ns}$ with spatial dimensions corresponding to the source $\mathbf{x}_{s}$ and receiver $\mathbf{x}_{r}$ coordinates, respectively, and wavefield registered at time $t$. For the general 3D case, a 5D volume with coordinates $(t, x_{r}, y_{r}, x_{s}, y_{s})$ is used to store the recorded data instead. We restrict ourselves to the 2D survey, organize the observed data in the space-frequency domain $(\omega, x_{r}, x_{s})$, and consider wavefield interpolation along the spatial dimensions for each temporal frequency $\omega$. Hence, a frequency slice comprising $m$ receivers and $n$ sources is the matrix $\mathbf{X} \in \mathbb{R}^{m \times n}$ where the $\omega$ dependency is omitted for ease of notation.

\begin{figure}[!t]
    \centering
    \includegraphics[scale=0.4]{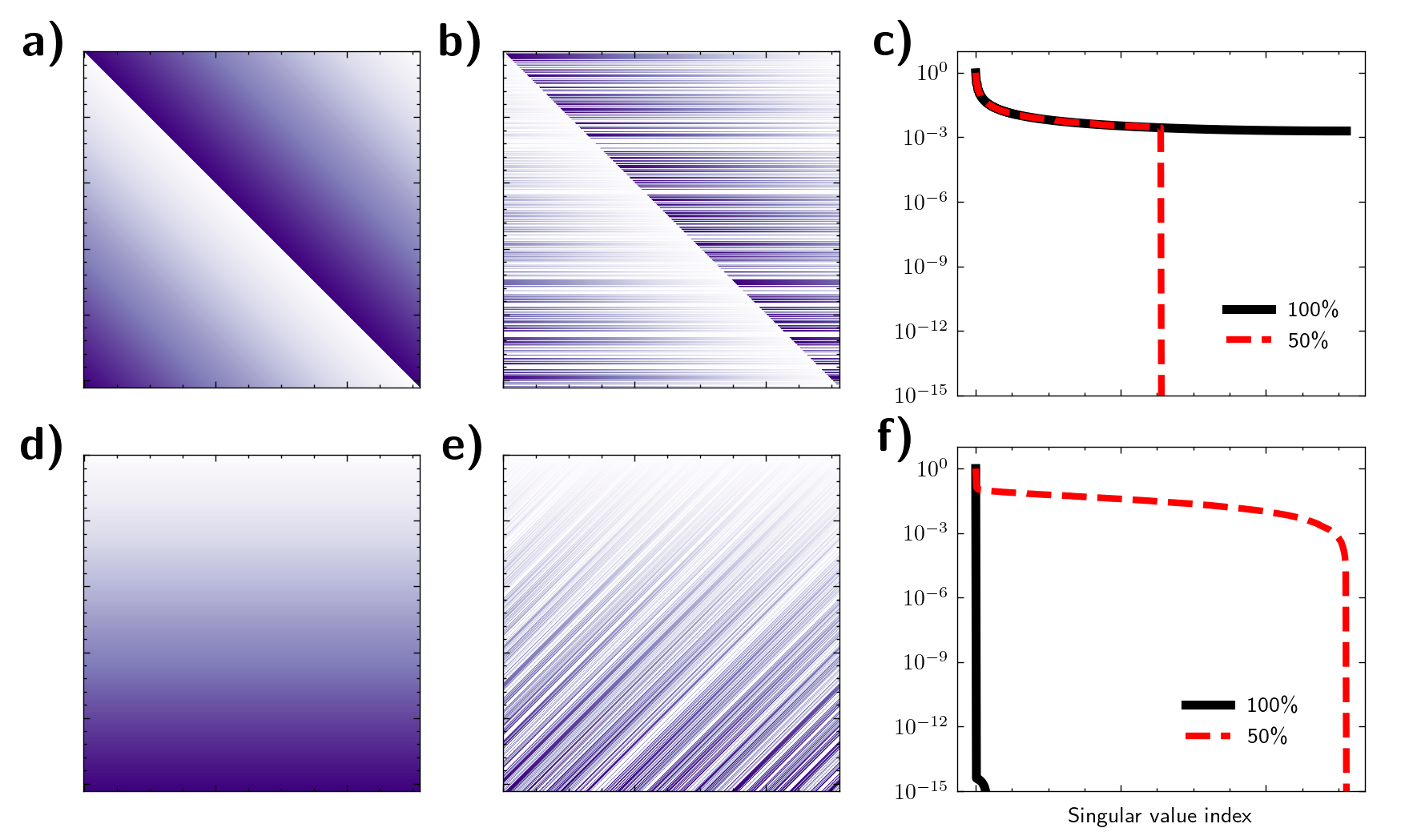}
    \caption{
    Rank modulation via cyclic-shear transformation. (a) The circulant matrix depicts a steady energy distribution along diagonal entries with a consistent variation across the antidiagonal direction. $50 \%$ Random row/column decimation (b) precipitates an abrupt singular value decay (dashed red line) in contrast to the original broadband spectrum (solid black line) (c). Likewise, the cyclic rearrangement induced by $\mathbf{S}^{\ast}_{y}$ turns the circulant into a low-dimensional array (d), (f) (solid black line), and makes the sampled array high-rank (e), (f) (dashed red line).
    }
    \label{fig_4}
\end{figure}

\subsection{Cyclic-shear preconditioner}\label{sec_3.1}
If the seismic survey presents with vacant receivers, records of the entire wavefield exhibit missing traces at such places for all sources ignited in a simultaneous or sequential mode. This situation translates into empty columns in the seismic volume. For this reason, instead of random missing entries, we want to complete $\mathbf{X}$ from data with randomly missing columns, rows, or even both, employing the same optimization procedure. Under such sampling conditions, any low-rank matrix algorithm fails to recover the data since the zero entries must be sufficiently scattered throughout the matrix \cite{Candes2010A, Candes2012}. We confront the adverse sampling protocol by introducing a transformation $\mathcal{S}: \mathbb{R}^{mn} \to \mathbb{R}^{m n}$ that preserves the random column selection in $\mathbf{X}$ and ensures probing any row and column at least once in the transform domain:
\begin{align*}
    [\mathcal{S}_{x}(\mathbf{X})]_{i, j}  \: &= \: \mathbf{X}_{\pi_{k}(i), j} \:,
    \label{eq12.1}\tag{12.1}     \\ 
    [\mathcal{S}_{y}(\mathbf{X})]_{i, j}  \: &= \: \mathbf{X}_{i, \pi_{k}(j)} \:,
    \label{eq12.2}\tag{12.2}  
\end{align*}
with subscripts $x$ and $y$ specifying the axis along which the cyclic permutation $\pi$, as stated in \textit{Definition} \ref{def_1}, is taken. We omit the subscripts whenever the mathematical statements apply to both cases to simplify the notation. Here, the shifting quantity $k$ is a free parameter which can be readily set among columns/rows. In particular, the corresponding matrix representation $\mathbf{S} \in \mathbb{R}^{mn \times mn}$ is written in block diagonal form for both permutation axis
\begin{align*}
    \mathbf{S}_{x}  \: &= \: \mbox{diag}(\mathbf{\Pi}_{1}^{k}, \mathbf{\Pi}_{2}^{k}, ..., \mathbf{\Pi}_{m}^{k})     \: , \\
    \mathbf{S}_{y}  \: &= \: \mbox{diag}(\mathbf{\Pi}_{1}^{k}, \mathbf{\Pi}_{2}^{k}, ..., \mathbf{\Pi}_{n}^{k})     \: , 
\end{align*}
where the building blocks $\mathbf{\Pi}_{1}^{k}, \mathbf{\Pi}_{2}^{k}, ..., \mathbf{\Pi}_{m}^{k} \in \mathbb{R}^{n \times n}$, and $\mathbf{\Pi}_{1}^{k}, \mathbf{\Pi}_{2}^{k}, ..., \mathbf{\Pi}_{n}^{k} \in \mathbb{R}^{m \times m}$ are independent cyclic shift matrices, as introduced in the \textit{Definition} \ref{def_2}. This operator inherits the property $\mathbf{S}^{\ast} = \mathbf{S}^{-1}$, and the composition rule $\mathbf{S}_{xy} = \mathbf{S}_{x} \mathbf{S}_{y}$ leads to a simultaneous row-column shift when applied on $\mathbf{X}$. 

The operation $\mathcal{S}^{\ast}(\mathbf{X}) = \mathbf{S}^{\ast} \mbox{vec}(\mathbf{X})$ results in a surrogate of the original matrix carrying the same information and can potentially recover the conditions for matrix completion when serving as a preconditioner in the problem \eqref{eq4}. Let $\mathbf{S}$ be a non-singular matrix, instead of solving the problem \eqref{eq4}, let us solve the preconditioned system:
\begin{equation}
    \min_{\mathbf{z}} \: \| \mathbf{R} \mathbf{S} \mathbf{z} - \mathbf{y} \|_{2}^{2} \:+\: \mu \| \mathbf{Z} \|_{*} \:.
    \label{eq13}\tag{13}
\end{equation}
We now expect to recover a mapping $\mathbf{Z}$ of the original matrix for which the sampling regime conforms with a low-coherence matrix in the new domain. In simple terms, the mapping $\mathcal{S}$ breaks the coherence of the missing data structure in the column/row spaces under the standard basis. Once $\mathbf{Z}$ is completed, the information is rearranged back through the forward operation $\mathbf{x} = \mathbf{S} \mathbf{z}$. This thesis proves effective when $k$ is set as a function of the column/row index so that the entries $X_{ij}$ transform under a cyclic shear mapping, i.e., $k(i) = \lfloor \alpha i \rfloor$ for rows and $k(j) = \lfloor \beta j \rfloor$ for columns, with $\lfloor \: \rfloor$ denoting the floor function. Given this argument, parameters $\alpha$ and $\beta$ effectively determine the degree of rank reduction and turn the operation into a form of rank modulation. Provided $k$ follows this constraint, we call $\mathbf{S}$ the \textit{cyclic-shear preconditioner}. 
    
\begin{figure}[!b]
    \centering
    \includegraphics[scale=0.47]{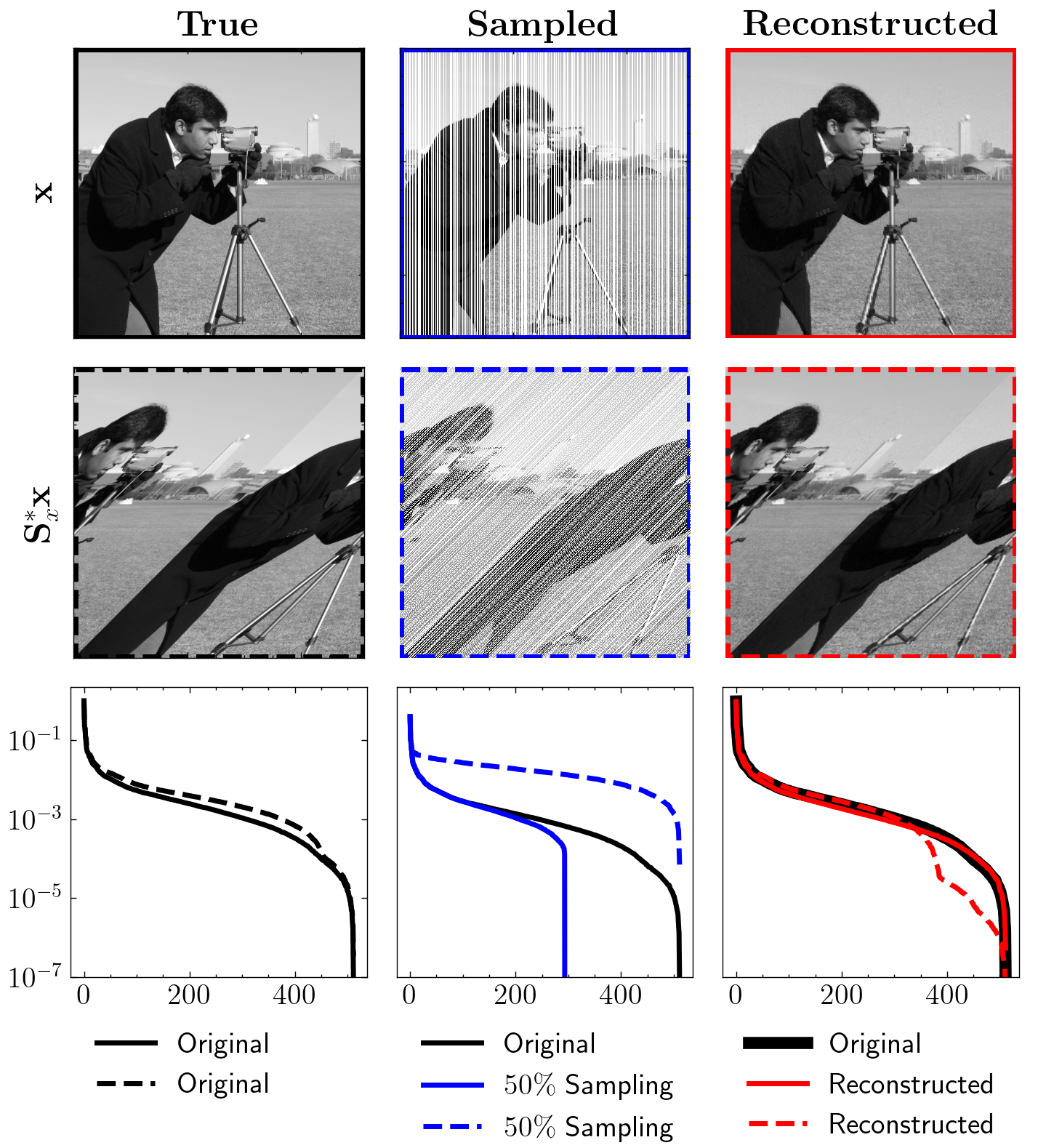}
    \caption{
    Digital image reconstruction via LRMC with $50\%$ randomly missing columns. The recovered image achieves a signal-to-noise ratio of $24.5 \: \text{dB}$. All ambient images in the first-panel row lie in the original domain, whereas, in the second, all panels belong to the codomain of the operator $\mathbf{S}^{\ast}_{x}$. The last panel row shows singular value spectra for all images above with colour and line styles referring to the frame enclosing the images. Empty columns raise the singular values in the transformed domain contrary to the sudden drop observed in the original domain (blue lines).
    }
    \label{fig_5}
\end{figure}

\begin{figure}[!ht]
    \centering
    \includegraphics[scale=0.075]{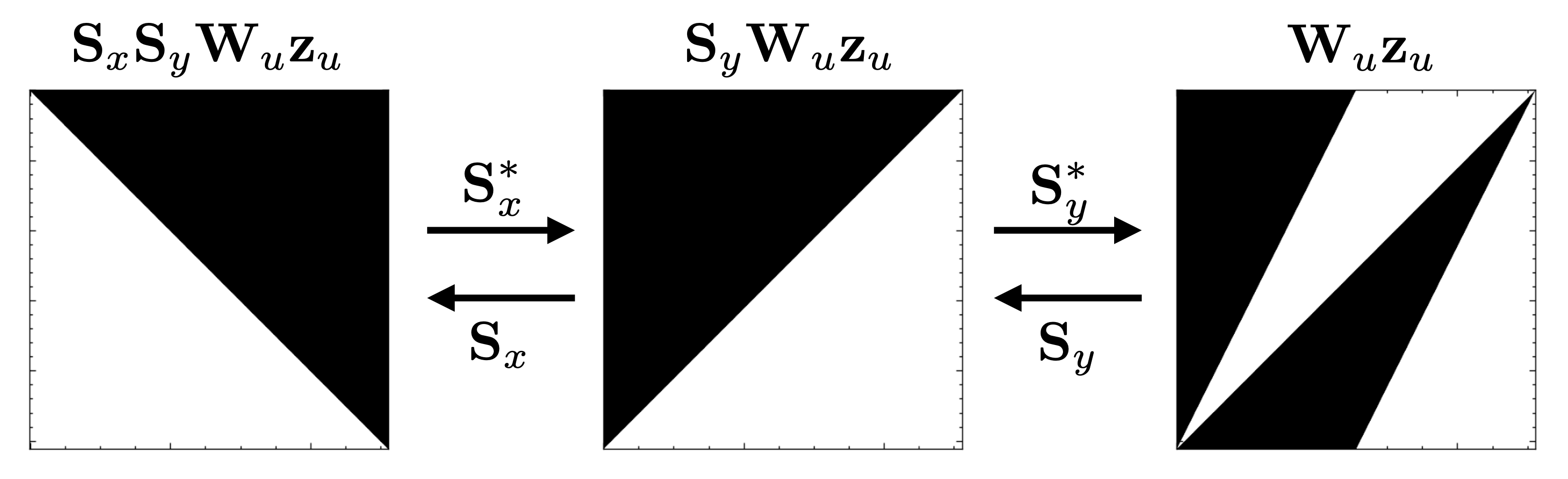}
    \caption{
    The splitting operator $\mathbf{W}_{u}$ reshapes the auxiliary model vector $\mathbf{z}_{u}$ into its matrix form and extracts the upper patch before flattening it back into a vector form. This operation is chained with the shear operation $\mathbf{S}_{x}\mathbf{S}_{y}$, taking the model from a reduced rank domain into the regular high-dimensional domain through a row-column cyclic shift. In essence, the composition rule maps the matrix's superdiagonal into two distinct blocks, as shown in the rightmost panel.
    }
    \label{fig_6}
\end{figure}

To illustrate the idea behind the cyclic-shear transformation, we analyze its effect on the circulant matrix $\mathbf{C} \in \mathbb{R}^{n \times n}$ \cite{Ingleton1956, Chan1991}. Using modular notation, the circulant matrix is defined as $C_{ij} = c_{(j - i) \: \text{mod} \: n}$. Equivalently, it can also be defined as a linear combination of $n$ cyclic shifts (\textit{Definition} \ref{def_2}) of the form
\begin{align}
    \mathbf{C} = c_{1} \mathbf{\Pi}^{0} + c_{2} \mathbf{\Pi}^{1} + \cdots + c_{n} \mathbf{\Pi}^{n-1} \: .
    \label{eq14}\tag{14}
\end{align}
We refer to it as the associated matrix polynomial of degree $n-1$. Vectorizing the circulant matrix into a long one-dimensional vector and applying $\mathcal{S}^{\ast}$ reduces to the rank-$1$ matrix $\mathbf{c}_{n} \otimes \mathbf{1}$ (see Appendix \ref{append_1} for details), where $\mathbf{1} = [1, 1, ..., 1]^{T} \in \mathbb{R}^{n}$ is the all-ones vector, $\mathbf{c}_{n} = [c_{1}, c_{2}, ..., c_{n}]^{T} \in \mathbb{R}^{n}$ is a generic vector, and $\otimes$ denotes the outer product. Figures \ref{fig_3} and \ref{fig_4} show an example of a circulant matrix and the resulting rank-$1$ matrix under the operator $\mathbf{S}_{y}$ and its adjoint. The matrix $\mathbf{C}$, as shown in Figure \ref{fig_4}a, is high rank with slowly decaying singular values (Figure \ref{fig_4}c). Still, under $\mathbf{S}^{\ast}_{y}$, the circulant results in the low-rank analogue matrix $\mathbf{c}_{n} \otimes \mathbf{1}$ (Figure \ref{fig_4}d). More specifically, the cyclic shear promotes low dimensionality, meaning that the singular values quickly decay in the transform domain (Figure \ref{fig_4}f). An important observation is that column/row removal decreases the matrix rank as singular values are set to zero (Figure \ref{fig_4}b-c). Nevertheless, under $\mathcal{S}^{\ast}$, this scheme translates into a general lifting of singular values making the sampled matrix high rank (Figure \ref{fig_4}e-f). The previous statement motivates matrix completion in the codomain of $\mathcal{S}^{\ast}$, where the target matrix now presents sufficiently incoherent measurements.

Back to our image completion example, we adapt the restriction operator $\mathbf{R}$ to randomly discard 50$\%$ of the columns from the $n \times n$ benchmark matrix to form the decimated vector $\mathbf{y}$, define the cyclic-shear $\mathbf{S}_{x}$ through $k(i) = i$, and minimize the preconditioned program \eqref{eq13}. Our test (Figure \ref{fig_5}) confirms a thriving quality in the recovered image, whose overall performance is competitive with the one observed previously (Figure \ref{fig_2}). In effect, a column-wise inspection in the transform sampled panel (Figure \ref{fig_5}) uncovers random missing entries as a consequence of the structure-revealing preconditioner transforming the cartesian sampling mask into a cyclic shear mask that closely resembles the ideal random sampling scheme.

\subsection{Seismic reciprocity}\label{sec_3.2}
In many instances, the principle of seismic source-receiver reciprocity serves as a physics-driven approach for building strong additional prior information in inverse problems where the symmetry properties of the sought-after wavefield can be used to constraint the optimization process by enforcing the subsurface response to remain intact when sources and receivers are interchanged in a common grid for the interpolated wavefield \cite{VderNeut2017, Kumar2020, Vargas2022}. We formally restrict the inverse problem by introducing an additional preconditioner such that
\begin{equation}
    \min_{\mathbf{z}} \: \| \mathbf{R} \mathbf{\Upsilon} \mathbf{S} \mathbf{z} - \mathbf{y} \|_{2}^{2} \:+\: \mu \| \mathbf{Z} \|_{*} \:
    \label{eq15}\tag{15}
\end{equation}
results in a completed wavefield $\mathbf{X}$ that is invariant when the spatial coordinates are transposed. In an iterative scheme, the operator $\mathbf{\Upsilon} = 1/2 (\mathbf{I} + \mathcal{T})$ reshapes the model vector into a matrix and takes the average between the current solution at a given iteration and its transposed version before reshaping the result back to a one-dimensional vector. Note that $\mathcal{T}$ is the transpose operator, and $\mathbf{I}$ the identity matrix. This physical prior produces stable results by discarding any solution for which the principle of reciprocity is not satisfied. In particular, the constraint aids matrix completion in cases when the survey presents disjoint sources and receivers or those in which one of the directions is densely sampled, but the other is not.
 
\subsection{Matrix splitting}\label{sec_3.3}
\begin{figure*}[!b]
    \centering
    \includegraphics[scale=0.41]{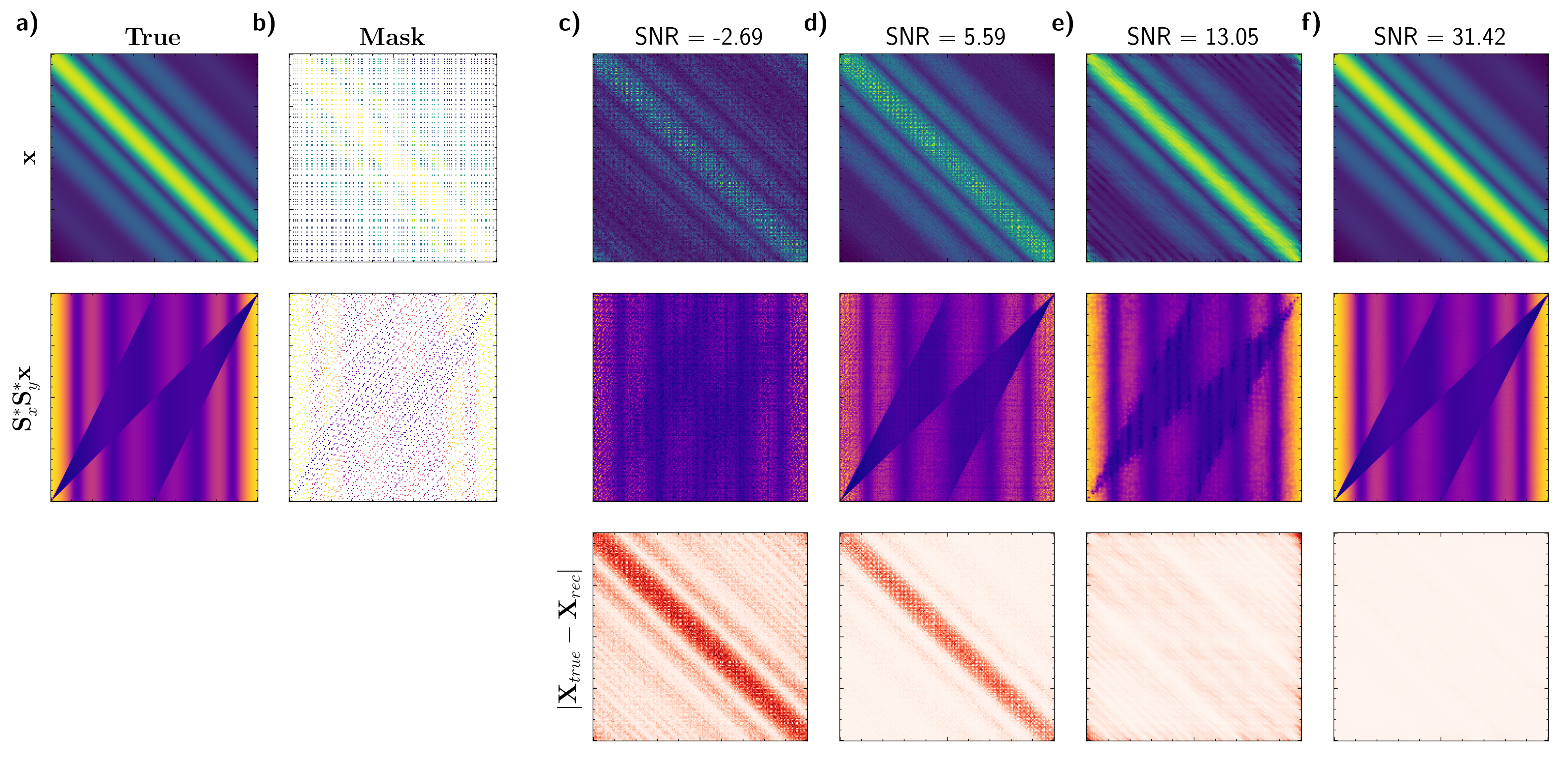}
    \caption{
    Frequency-slice extracted at $15$Hz from the 2D layered synthetic reflection data for reconstruction. From top to bottom monochromatic slices are sorted in the $x_r$-$x_s$ domain (first-panel row), cyclically sheared across both source and receiver axes (second-panel row), and rearranged back to show the corresponding residual error after matrix completion (last-panel row). (a) The fully sampled slice for recovery evaluation of the (b) given $90\%$ masked data. (c) The partially completed matrix using program \eqref{eq13} is (d) improved by enforcing reciprocity and matrix splitting priors as in Equation \eqref{eq16}; yet, a significant enhancement is revealed when proximity similarity (Equation \eqref{eq17}) is set off in both (e) non-reciprocal and (f) reciprocal solutions.   
    }
    \label{fig_7}
\end{figure*}

Even though the cyclic-shear preconditioner $\mathbf{S}$ proves effective for simultaneous source-receiver seismic data reconstruction, the geometrical nature of this transformation results in a sharp data transition as observed in Figure \ref{fig_7}a, where a high-rank fully sampled monochromatic frequency slice (first row) is cyclically shifted to form its low-rank counterpart (second row). Note how honed distinct segments arise inside the matrix, breaking the desired smoothness across adjacent values. Under such conditions, one can expect the nuclear norm penalty to smear out any sudden variation during the optimization process. See, for instance, the reconstruction result in Figure \ref{fig_7}e. In general, one cannot make sure data laying towards a specific edge share a degree of similarity with the opposite side; however, $\mathbf{S}$ brings boundary data closer, breaking the natural correlation among neighbouring entries. As expected, low-rank completion algorithms cannot discriminate the artificial affinity induced among neighbouring entries initially allocated on the edges. As a result, the optimization program will tend to align contiguous data near the sharp discontinuity, see Figure \ref{fig_7}e. Fortunately, this effect is well localized, affecting only edge data. On that account, we introduce operators $\mathbf{W}_{l}$ and $\mathbf{W}_{u}$ to decompose the seismic frequency slice into lower and upper windows (Figure \ref{fig_6}) such that $\mathbf{x} = \mathbf{\Upsilon} \mathbf{S} [\mathbf{W}_{l}\mathbf{z}_{l} + \mathbf{W}_{u}\mathbf{z}_{u}]$, with $\mathbf{z}_{l}$, $\mathbf{z}_{u}$, auxiliary disjoint matrices. The nuclear norm penalized problem is then extended to correct smeared edge effects,
\begin{equation}
    \min_{\mathbf{z}} \: \| \mathbf{R} \mathbf{\Upsilon} \mathbf{S} \mathbf{W} \mathbf{z}  - \mathbf{y} \|_{2}^{2} \:+\: \mu \sum_{i\in{l, u}} \| \mathbf{D}_{i} \mathbf{z} \|_{*} \:,
    \label{eq16}\tag{16}
\end{equation}
with $\mathbf{D}_{i}$ an operator such that $\mathbf{z}_{l} = \mathbf{D}_{l} \mathbf{z}$, and $\mathbf{z}_{u} = \mathbf{D}_{u} \mathbf{z}$. This formulation solves for $\mathbf{z} = [\mathbf{z}_{l} \; \mathbf{z}_{u}]^{T}$ and uses the windowing operator $\mathbf{W} = [\mathbf{W}_{l} \; \mathbf{W}_{u}]$ to extract the upper and lower independent parts of the solution forming an edge-corrected model vector $\mathbf{x} = \mathbf{\Upsilon} \mathbf{S} \mathbf{W} \mathbf{z}$. In ambient image reconstruction, this edge effect can go unnoticed by an observer's perception; however, working with seismic data may lead to wrap-around effects observed in the gathers at far offsets. In our experience, the correction can sometimes be dropped depending on how energy distributes across the entire domain. In particular, if most of the energy is allocated along the diagonal, i.e., zero-offset data, and it rapidly decays towards the edges, one can disregard the windowing operator and still recover a high-quality signal. Conceptually, seismic data tends to arrange this way on the grounds of geometrical spreading, an argument exploited in low-rank compression of reflection data \cite{Hong2021, Ravasi2022}. 

\subsection{Proximity similarity regularizer}\label{sec_3.4}
Standard prestack seismic volumes organize recordings according to coordinate systems, promoting neighbouring observations to remain closer to one another. This arrangement promotes a level of similarity among contiguous receiver/source gathers and serves as an additional constraint imposed by wave physics. We leverage the local similarity properties of the seismic wavefield and propose the constrained minimization algorithm:
\begin{equation}
    \min_{\mathbf{z}} \: \| \mathbf{R} \mathbf{\Upsilon} \mathbf{S} \mathbf{z} - \mathbf{y} \|_{2}^{2} \:+\: \mu_{1} \| \mathbf{Z} \|_{*} \:+\: \mu_{2} \| \mathbf{\nabla} \mathbf{z} \|^{2}_{2} \:,
    \label{eq17}\tag{17}
\end{equation}
for cases where edge effects are negligible, and
\begin{equation}
    \min_{\mathbf{z}} \: \| \mathbf{R} \mathbf{\Upsilon} \mathbf{S} \mathbf{W} \mathbf{z} - \mathbf{y} \|_{2}^{2} \:+\: \mu_{1} \sum_{i} \| \mathbf{D}_{i} \mathbf{z} \|_{*} \:+\: \mu_{2} \sum_{i} \| \mathbf{\nabla} \mathbf{D}_{i} \mathbf{z} \|^{2}_{2} \:,
    \label{eq18}\tag{18}
\end{equation}
for those leading to wrap-around effects induced by nuclear norm edge smearing. Note that, the auxiliary penalty term enforces proximity similarity across consecutive locations by minimizing the first spatial derivative of the solution in the L2 sense, i.e., it penalizes a measure of the model roughness \cite{Hansen2010}. While the solution's low-rank aspect is mathematical in nature and seeks structure in the alignment of the column and row spaces, the lateral similarity prior is physics-inspired and yields smooth transitions deprived of sharp edges. This regularization strategy renders an enhanced reconstruction by complementing the action of the nuclear norm while adhering to the physical features characteristic of wavefield-based data. In addition, we observe an acceleration in the algorithm's convergence rate that significantly reduces the number of iterations needed to deliver superior results, which is indicative of the desired numerical effect of preconditioning.    

\begin{figure*}[!t]
    \centering
    \includegraphics[scale=0.45]{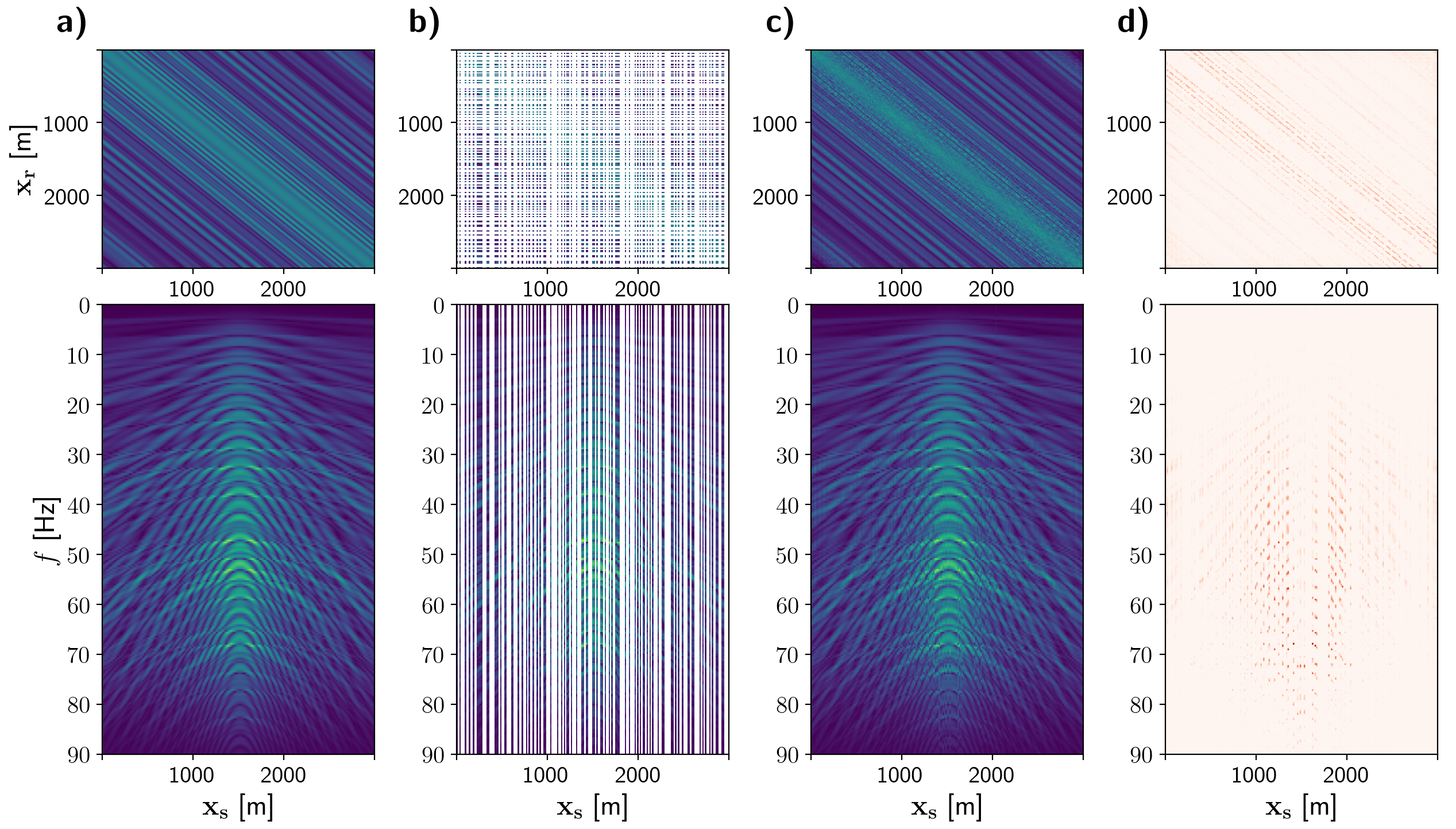}
    \caption{
    Frequency-space domain interpolation of synthetic data in a stratified earth model. (a-d) frequency slices at $60$ Hz (top panel row), and $f$-$x$ common receiver gathers (bottom panel row). (a) The Fully sampled data volume is (b) decimated up to  $80\%$ missing traces with $55\%$ empty rows (receivers) and $55\%$ empty columns (sources). (c) The reconstructed volume using seismic reciprocity, matrix splitting, and proximity similarity priors as described in the proposed algorithm (Equation \eqref{eq17}) with a final SNR $18.25$ dB. (d) Residuals between the (a) benchmark and (e) the recovered data.
    }
    \label{fig_8}
\end{figure*}

\section{Numerical experiments}\label{sec_4}
We consider several tests on synthetic and field data to evaluate the performance of nuclear norm minimization for interpolating 2D seismic lines. Our numerical test focuses on the simultaneous reconstruction of missing sources and receivers, whereas interpolation along a single axis is regarded as a particular case based on the survey design characteristics. Our purpose is to infer the absent energy at random locations that constitute the vacant traces as accurately as possible and illustrate the preconditioned-based program's effectiveness in various 2D geological settings. The first experiment examines the synthetic broadband response from a 1D layered medial that induces a wide range of dips and symmetrically reflects energy in all directions. Our next target considers a more complex scenario mimicking a sedimentary basin extracted from the SEAM Phase I model \cite{Fehler2011}. We conclude with a field data example illustrating source-receiver reconstruction on a seismic line from the Gulf of Suez \cite{Abouelela2020}.  

To evaluate the performance of the proposed formulation, we compare all reconstructed wavefields against the corresponding alias-free seismic response using the following signal-to-noise ratio
\begin{align*}
    SNR = 10 \log_{10} \frac{\| \mathbf{X}_{true} \|^{2}_{F}}{\| \mathbf{X}_{true} - \mathbf{X}_{rec} \|^{2}_{F}} \: ,
\end{align*}
where $\mathbf{X}_{true}$ is the true fully-sampled data and $\mathbf{X}_{rec}$ is the interpolated data.
\begin{figure*}[!t]
    \centering
    \includegraphics[scale=0.45]{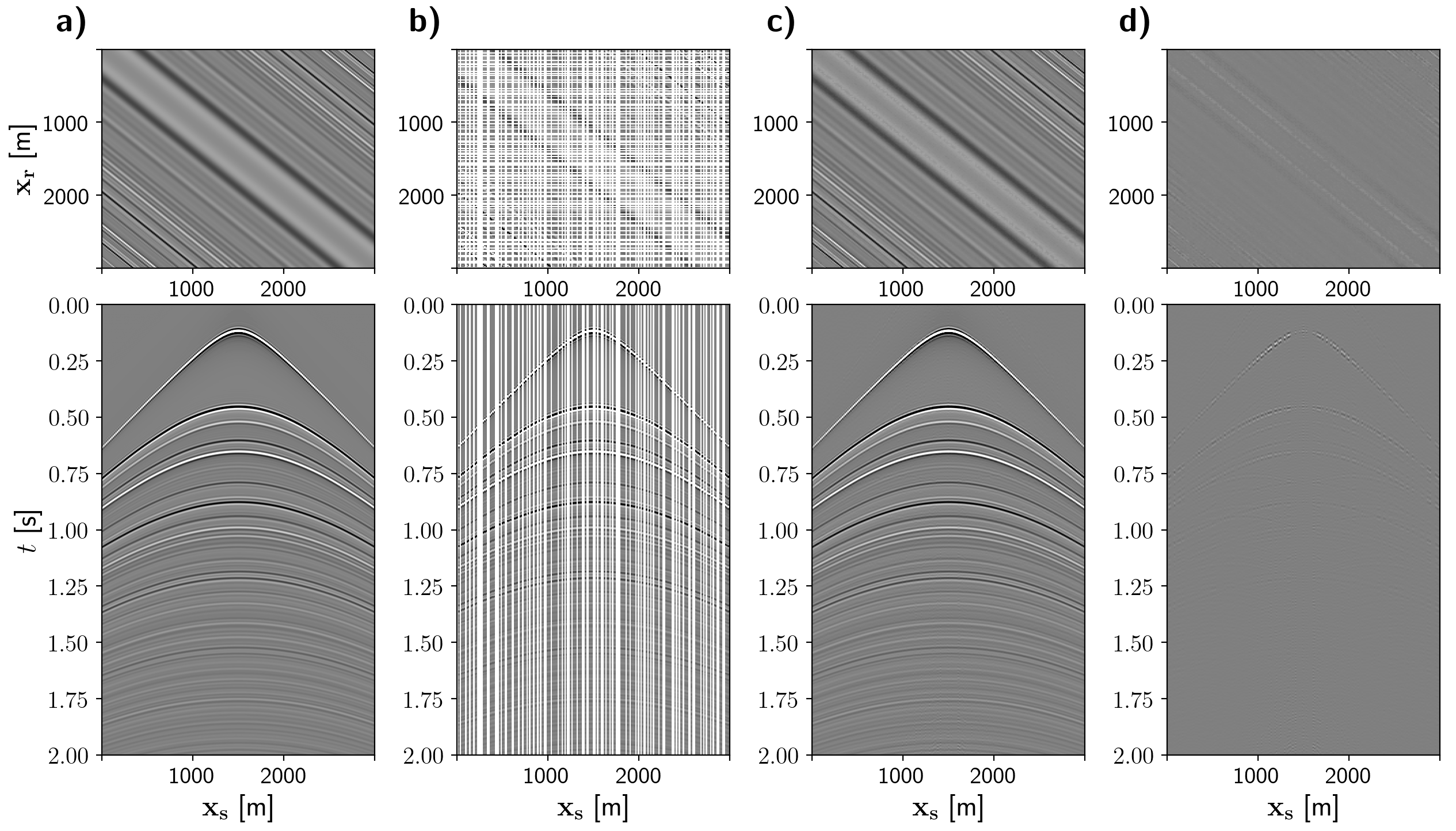}
    \caption{
     Synthetic data set in a layered 2D earth model. (a-d) Time slices at $1.2$ s (top panel row), and time-space common receiver gathers (bottom panel row). (a) Modelled initial data. (b) Decimated data with $80\%$ missing traces resulting from $55\%$ receivers removal along with $55\%$ absent sources. (c) Simultaneous reconstruction of sources and receivers using the proposed rank-constrained algorithm with cyclic shear, reciprocity and lateral smoothness constraints (SNR $18.25$ dB). (d) The residual between the (c) reconstruction and the (a) reference data.
    }
    \label{fig_9}
\end{figure*}

\begin{figure*}[!b]
    \centering
    \includegraphics[scale=0.297]{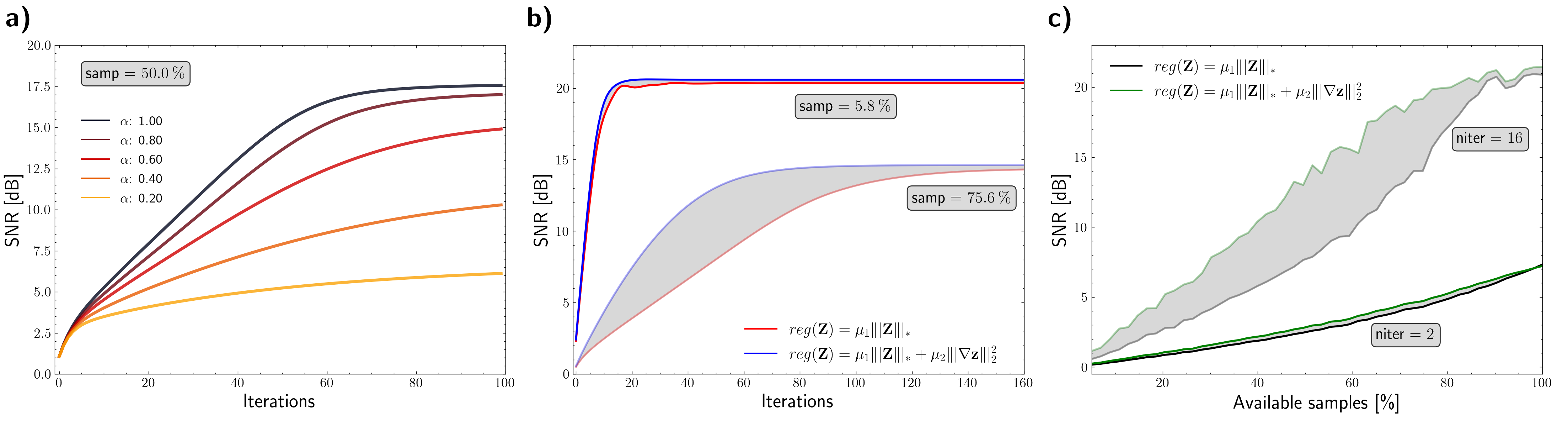}
    \caption{
    Peak performance of cyclic shear preconditioner and proximity similarity regularizer. (a) Signal-to-noise ratio as a function of iterations for multiple cyclic-shear slopes $\alpha$. (b) The Laterally constrained inversion ensures fewer iterations are required as the sampling rate increases (solid-blue line), opposite to classic nuclear norm minimization (solid-red line), which converges at a slower pace. (c) The same behaviour is observed for a fixed number of iterations and varying percentages of available samples. Lateral constraints (solid-green line) result in higher SNR than conventional LRMC (solid-black line) at early iterations.
    }
    \label{fig_10}
\end{figure*}

\subsection{Synthetic data - Stratified medium}\label{sec_4.1}
We first consider the problem of reconstructing a noiseless synthetic reflection response produced by a 2D stratified earth model consisting of twelve flat layers. The model extends $4.0$ Km in the horizontal direction, comprising $2.0$ Km depth, and is designed to generate both short and long-period internal multiples exhibiting a wide range of plane waves with different dipping angles. In particular, the 2D layered model smoothly spreads scattered energy in the mid to large offsets, creating a balanced signal distribution across all frequencies and constitutes a simple reference for matrix completion. Using a broadband impulse source with a flat spectrum in the range of $1$–$80$ Hz, an acoustic solver simulates $201$ sources sequentially injected on the model surface; then, the medium response is recorded with a $15$ m receiver interval network. The temporal length of the gathers is $2.7$ s with a sampling rate of $2.5$ ms, which leads to a full dataset containing $201 \times 201 = 40401$ seismic traces with $1080$ time samples in each trace. We decimate the seismic volume through the restriction operator $\mathcal{R}$ before moving the data into the space-frequency domain, where the low-dimension properties are expected to be revealed. 

\begin{figure*}[!t]
    \centering
    \includegraphics[scale=0.45]{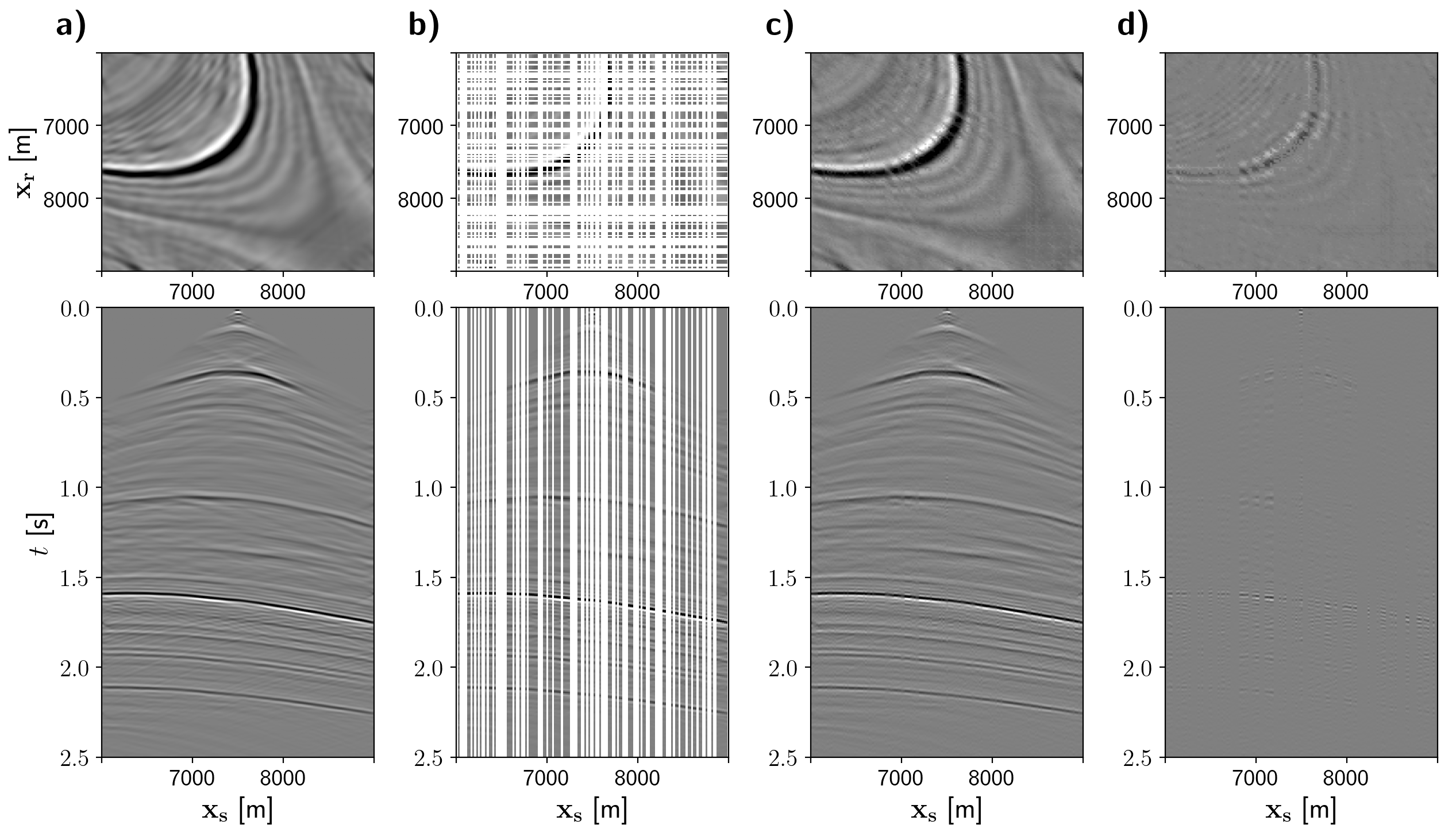}
    \caption{
    Synthetic data set in a section of the 2D SEAM model. (a) Numerically modelled data. (b) Sub-sampled data with absent sources and receivers resulting in $75\%$ missing traces. (c) Simultaneous reconstruction of sources and receivers with local similarity constraints and reciprocity prior (SNR $9.6$ dB). (c) Reconstruction error. Time slices (top panel row) are extracted at $1.6$ s.
    }
    \label{fig_11}
\end{figure*} 

To uncover the benefits of adopting the proposed chained preconditioner while enforcing proximity similarity among neighbouring traces, we conduct seismic interpolation on a low-frequency monochromatic slice extracted from the data at $15$ Hz. This test comprises a jittered sampling scheme \cite{Hennenfent2008} that independently removes $70\%$ of the sources and $70\%$ of the receivers, resulting in a problem with $90\%$ missing data. Figure \ref{fig_7} presents the reconstruction results of various regularization and preconditioning strategies. It is evident that our cyclic-shear transformation $\mathbf{S}_{x}\mathbf{S}_{y}$ successfully reveals the low-rank character of seismic data and enables data recovery by promoting sparsity in the singular values spectrum.

Figure \ref{fig_7}a shows the original frequency slice followed by its masked version (Figure \ref{fig_7}b) in the standard $x_r$-$x_s$ domain (first-row panel) and in the rank-revealing space (second-row panel) induced by the cyclic shear map. The first case is that of traditional matrix completion as stated in the problem \eqref{eq13} and shown in Figure \ref{fig_7}c. Despite the substantial column/row gap filling, a significantly low signal-to-noise ratio (SNR $-2.69$ dB) exposes the limitations of raw low-rank matrix completion in correctly reconstructing low-frequency data without additional constraints, even in a reduced-rank domain. In the second case (Figure \ref{fig_7}d), we endow the problem with additional reciprocity and matrix-splitting priors to lift the recovery quality to SNR $5.59$ dB. Still, it is a relatively low performance; however, invoking explicit lateral proximity similarity among neighbouring sources and receivers discloses the full potential of the proposed algorithm without having to leverage the low sampling rate. Figure \ref{fig_7}e depicts a significant image enhancement with SNR $13.05$ dB when such regularizer acts in combination with the nuclear norm, i.e., $reg(\mathbf{Z}) = \mu_{1} \| \mathbf{Z} \|_{*} \:+\: \mu_{2} \| \mathbf{\nabla} \mathbf{z} \|^{2}_{2}$ in the problem \eqref{eq13}. By extension, one can constrain the solution to obey reciprocity as in the program \eqref{eq17} while controlling edge effects via matrix splitting. In this case, the restored array (Figure \ref{fig_7}f) displays superior results with a remarkable SNR $31.42$ dB.

Solving the simultaneous reconstruction problem \eqref{eq17} on a frequency-by-frequency basis with $80\%$ jittered subsampled data, $55\%$ missing sources together with $55\%$ missing receivers, derives in the 3D frequency-space volume shown in Figure \ref{fig_8}. For reference, Figure \ref{fig_8}a shows the gap-free data on a regular grid, while in Figure \ref{fig_8}b, the masked data are presented. Figure \ref{fig_8}c shows the restored $x$-$f$ spectrum followed by the residual error in Figure \ref{fig_8}d. At this point, we move the data to the space-time domain to verify the retrieved solution quality. For consistency, Figure \ref{fig_9} follows the same panel displaying format as in Figure \ref{fig_8}. Inspection of the restored data (Figure \ref{fig_9}c) evidence a reciprocal signal that follows a smooth lateral transition in the vicinity of the filled gaps, adhering to the local similarity constraint. The residual (Figure \ref{fig_9}d) is barely visible, and the reconstruction quality marks SNR $18.25$ dB. 

\subsection{Performance of lateral similarity constraint}\label{sec_4.2}
A detailed analysis of the influence of the proposed low-rank optimization priors is a prerequisite to establishing the role of preconditioning, sampling rates, solver iterations and regularization strategies. It is well known that reciprocity accelerates the algorithm's convergence rate; similarly, removing the edge effects whenever critical is simply achieved by invoking the matrix splitting operator $\mathbf{W}$. What is not obvious is how to set up the free parameters $\alpha$ and $\beta$ controlling the permutation indices $k(i) = \lfloor \alpha i \rfloor$ and $k(j) = \lfloor \beta j \rfloor$ in the cyclic shear preconditioner $\mathbf{S} = \mathbf{S}_{x}\mathbf{S}_{y}$. We conduct a series of matrix completion tests using various cyclic shear angles as a function of iterations. Figure \ref{fig_10}a shows a gradual increase in signal-to-noise ratio for slopes $0<\alpha<1$. If the matrix is square, the best choice is $\alpha = 1.0$; however, the optimal angle turns out to be $\alpha = m/n$ for a rectangular matrix. The same conclusion extends to $\beta$. In general, the guiding principle is to restrict the rank by aligning the matrix columns/rows through a cyclic shear that relocates empty columns/rows along the antidiagonal direction. 

The most conspicuous constraint imposed on the retrieved wavefields is the proximity similarity acting as an explicit regularizer. By evaluating its performance under multiple sampling rates as a function of solver iterations, the efficiency of such prior is subject to evaluation against that of the nuclear norm alone. Note that opposite to nuclear norm minimization, of mathematical character, lateral smoothness is motivated by wave physics; hence, they reinforce one other and contribute towards an enhanced solution using different mechanisms. Figure \ref{fig_10}b shows the reconstruction quality across iterations for a fully preconditioned nuclear norm minimization scheme (red-solid line) against its counterpart problem, which, in this case, is the option dressed with lateral constraints (blue-solid line). 

Suppose the input data are highly populated with almost no missing samples ($5.8 \%$ sampling rate). In that case, both programs quickly converge towards an accurate reconstruction. In contrast, both solutions instantly deviate when the sampling rate increases, resulting in a poorly recorded signal (Figure \ref{fig_10}b). In particular, for $75.6 \%$ missing samples, the laterally constrained program requires almost half fewer iterations to converge towards the proposed nuclear norm solution. Therefore, the new constraint increases the converse rate while enforcing wave physics on reconstructed signals. The same conclusion is drawn from Figure \ref{fig_10}c where the percentage of available samples ranges $0$-$100\%$, and the laterally constrained solution (solid-green line) depicts an overall higher signal-to-noise ratio in comparison to that of standard matrix completion (black-solid line). 
\begin{figure*}[!t]
    \centering
    \includegraphics[scale=0.45]{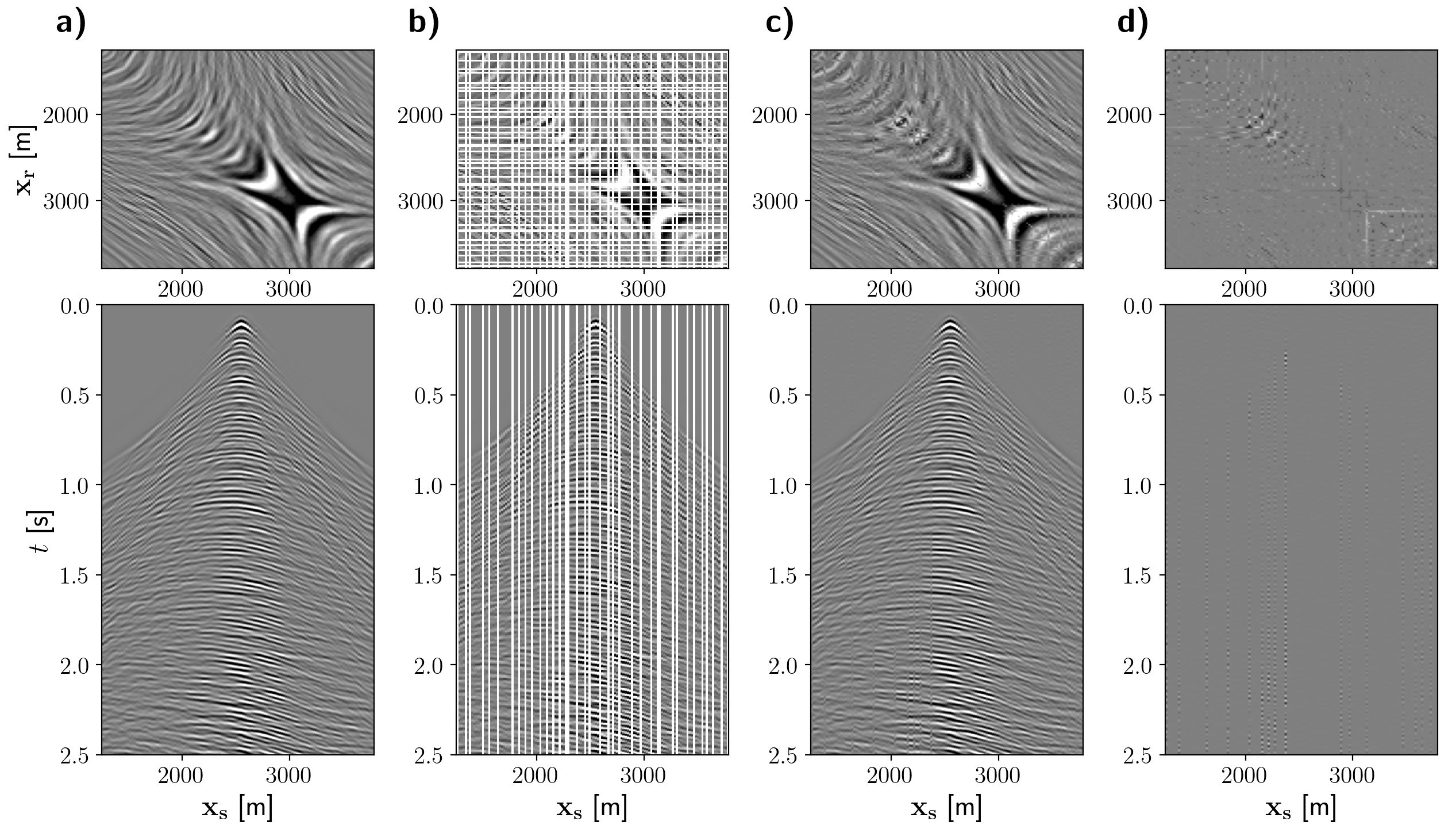}
    \caption{
     Simultaneous source-receiver interpolation of a seismic line from the Gulf of Suez. $50\%$ traces are removed from the seismic line, accounting for $30\%$ fewer sources besides $30\%$ missing receivers. (a) A Fully sampled common-receiver gather in the middle of the line, (b) the corresponding decimated gather with jittered shots removed, (c) the low-rank recovered counterpart (SNR $9.1$ dB), and (d) the residuals. (a-d) Time slices extracted at $1.85$ s (top panel row).    
    }
    \label{fig_12}
\end{figure*}

\subsection{Synthetic data - SEAM model section}\label{sec_4.3}
We now turn our attention to a synthetic data set modelled in a cross-section of the SEAM Phase I velocity model \cite{Fehler2011}. The selected region represents a seismically challenging earth model mimicking the natural geology of a layered sediment environment in the Gulf of Mexico with fine-scale stratigraphic features below seismic resolution that result in rich reflection dynamics exhibiting multiple wavefield scattering effects. The acquisition array involves $151$ co-located sources and receivers uniformly distributed every $20$ m over a $3$ km line. Data modelling is carried out with a broadband impulse-source wavelet in the rank $1$-$50$ Hz with maximum recording time $2.5$ s registered at $4$ ms sampling interval. The 3D array size we attempt to reconstruct is $625 \times 151 \times 151$ (Figure \ref{fig_11}). 

Data decimation is performed using a jittered sampling operator $\mathcal{R}$ acting along both source and receiver coordinates that retain the time axis intact and prevent large gaps in the target data. The interpolation procedure is carried out in the low-dimension domain induced by cyclically permuting the model parameter using the cyclic-shear mapping $\mathbf{S}_{x}\mathbf{S}_{y}$. The proposed optimization technique processes all sources and receivers through solver iterations, naturally exploiting the implicit data redundancy across the entire survey. Figure \ref{fig_11}a displays the reference reflection response with full source and receiver data next to its decimated version (Figure \ref{fig_11}b) with half of the sources and receivers removed, i.e., $75\%$ missing traces.

The reconstructed array and corresponding error panels are shown in Figures \ref{fig_11}c and \ref{fig_11}d. We use the primal-dual method with reinforced lateral constraints as indicated in Equation \eqref{eq17} and achieve SNR $9.6$ dB. In this case, the performance is boosted by demanding the solution to be reciprocal while enforcing sharp edges. Similarly, the residual panel (Figure \ref{fig_11}d) shows a relatively low amplitude that validates wavefield reconstruction in such a complex environment.

\begin{figure}[!t]
    \centering
    \includegraphics[scale=0.48]{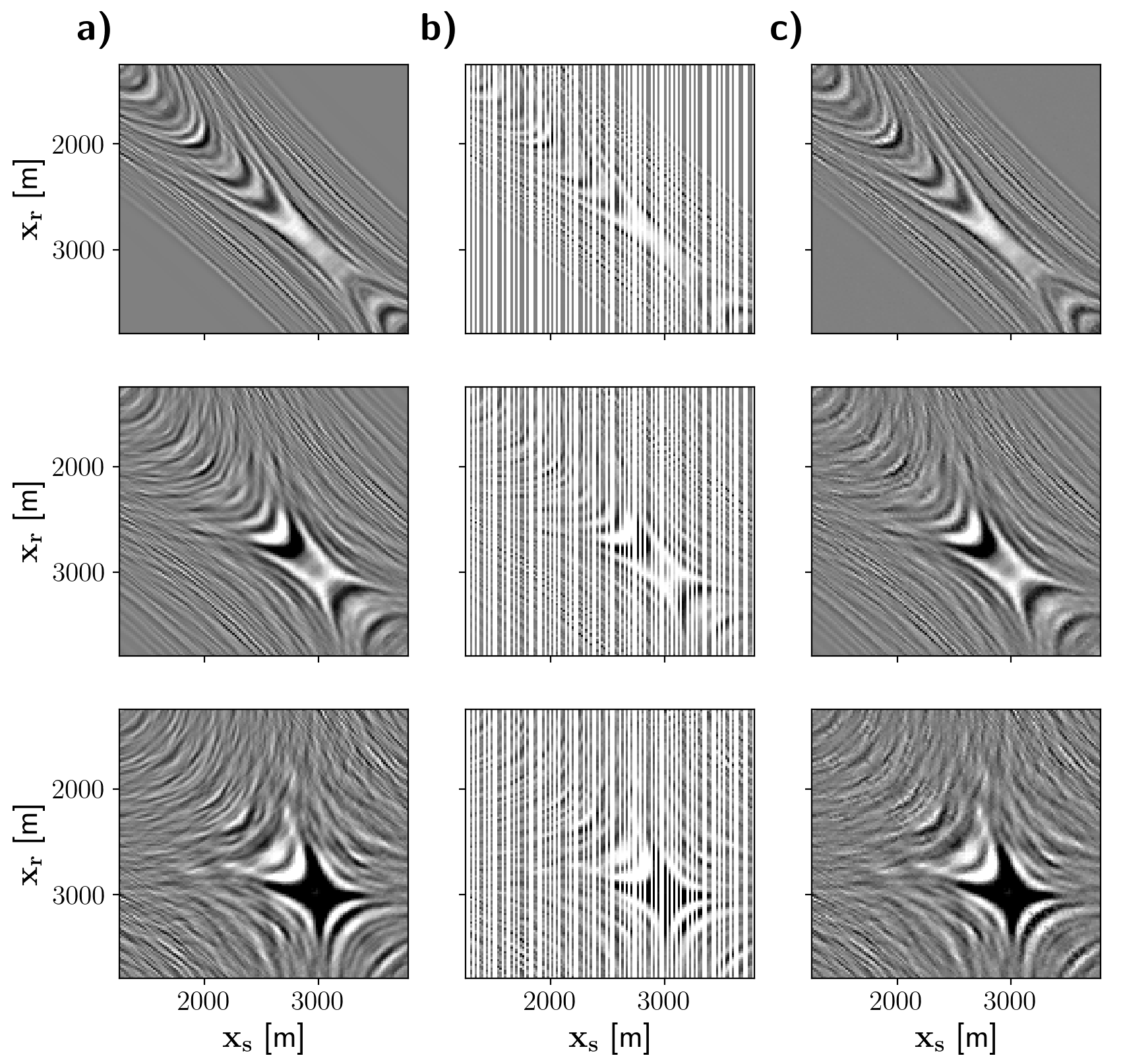}
    \caption{
     Low-rank recovery of the Gulf of Suez data set with $50\%$ missing shots. From top to bottom, time slices at $0.85$ s, $1.55$ s, and $2.35$ s are displayed. (a) Fully sampled data. (b) Jittered subsampled shots with full data on the receiver side. (c) Reconstructed wavefields with lateral constraints in a low-dimensional space (SNR $12.9$ dB).
    }
    \label{fig_13}
\end{figure}

\subsection{Field data - Gulf of Suez}\label{sec_4.4}
The Gulf of Suez constitutes a challenging region for seismic processing and imaging, partly due to the large impedance contrast inducing trapped multiples in the near-surface that manifest at late times but also because the narrow channel limits the acquisition of far offsets \cite{Abouelela2020}. Surrounded by such conditions, we conduct simultaneous seismic data interpolation in a seismic line from the Gulf of Suez and illustrate the effectiveness of low-rank recovery with lateral constraints for two sampling scenarios (Data publicly available at: \cite{Berg2013}). The reference dataset contains $128 \times 128$ sources and receivers distributed along $3.8$ Km every $20$ m with traces recorded at $0.005$ s sampling interval during $2.55$ s. Then, the total volume size is $510 \times 128 \times 128$.

First, simultaneous source-receiver reconstruction is carried out after removing $30\%$ jittered sources and receivers, accounting for a total $50\%$ trace decimation in the whole array. In Figure \ref{fig_12}, we display the central common-shot gather along with a time slice at $1.85$ s. Figure \ref{fig_12}a shows the reference data with all sources and receivers in place. In contrast, Figure \ref{fig_12}b exposes the extent of vacant grid points with respect to the reference. Upon inspection of Figure \ref{fig_12}c, it is evident that low-rank completion constrained by proximity similarity withstands the presence of noise and successfully reconstructs the missing information (SNR $9.1$ dB), leaving behind minimum residual energy (Figure \ref{fig_12}d).

Second, we consider the particular case of interpolating along the source side only (Figure \ref{fig_13}). This example assumes that all receivers keep recording during the acquisition experiment, whilst only $50\%$ jittered sources are ignited. This case is not different in that the algorithm displays high performance in reconstructing the missing sources; nonetheless, having access to a fully sampled axis better constrains the inversion, as verified by the higher SNR ($12.9$ dB).
    
\section{Discussion}\label{sec_5}
We present an efficient transformation $\mathcal{S}$ capable of revealing the seismic data's low-rank character, enabling matrix completion for seismic problems in an alternative domain. Contrary to the common practice of sorting the data in the midpoint-offset domain, which for large-scale problems may be an operation of significantly high computational cost, our strategy relies on a fast column/row-wise circular shift of the data array as stored in the regular acquisition domain. Even though the data array in the source-receiver domain is high rank as a consequence of strong diagonal entries (zero-offset energy) and subsequent off-diagonal oscillations, the new approach is designed to align the high energy contribution allocated along diagonal entries while preserving the characteristic local proximity among source-receiver pairs. Such transformation induces a modified domain where data naturally exhibit redundancy and discloses a strategy for rank modulation of matrices. It is worth mentioning that this operation is mathematically inspired and, therefore, deprived of the traditional physical meaning that midpoint-offset coordinates carry. Instead of transforming the whole data set from the source-receiver domain to the midpoint-offset coordinate system, the cyclic shear operation can be applied locally on a reduced data segment. This aspect opens new avenues for large-scale matrix completion problems implemented in parallel with the aid of High-Performance Computing facilities. Matrix completion in the cyclically sheared domain prevents the implicit zero-padding required when sorting the data in the midpoint-offset that results in data size duplication $mn \rightarrow (m+n)^{2}/2$; therefore, we expect that avoiding a $45$ deg rotation may help mitigate the gap between SVD-free techniques and nuclear norm minimization. Still, our strategy is not limited to nuclear norm regularized problems and can accommodate low-rank matrix factorization methods.

Despite the numerous theoretical benefits of leveraging the symmetry induced by seismic reciprocity, enforcing such a preconditioner poses different challenges when field data are under consideration. Modern Ocean Bottom Cable (OBC) or Node (OBN) receiver configurations are deployed at different depths, making it necessary to consider data redatuming as an additional operation to enable enforcing spatial reciprocity constraints. In the case of towed streamer marine and vibroseis land acquisition, a reciprocal solution requires simultaneous source/receiver interpolation to a grid of co-located sources and receivers, in addition to multicomponent wavefield considerations, which may be challenging to achieve. In spite of such practical challenges still to be overcome, the potential for a reciprocal reconstruction is clear. In particular, this constraint aids matrix completion in cases when the survey presents disjoint missing sources and receivers or those in which one of the axes is densely sampled but the other is not. We stress that getting more information from fewer samples is facilitated by reciprocity which generally forces the inversion to converge faster and promotes a robust solution in the presence of noise. 

A final aspect explored in this work concerns the introduction of additional constraints in the form of regularization. In particular, we highlight the benefits of enforcing explicit lateral similarity among contiguous sources and receivers. We observe an increase in the iterative solver's convergence rate when physical priors enforcing local closeness are incorporated into the objective function. In 3D seismic surveys, this can be achieved through distance-aware matrix reordering \cite{Ravasi2022} complemented with explicit lateral similarity as required in the problem \eqref{eq17} or \eqref{eq18}. Note that opposite to nuclear norm minimization, of mathematical character, controlling the degree of lateral roughness is motivated by wave physics; hence, they complement one another and contribute towards an enhanced solution using different mechanisms. More importantly, we showed how allowing such priors has the potential to recover highly decimated data beyond the limits required by low-rank matrix completion alone and could aid seismic reconstruction in complex environments where the wavefield typically undergoes a highly scattering propagation regime. 

\section*{Acknowledgments}\label{sec_6}
The authors express their gratitude to Haorui Peng, Leon Diekmann, Andreas Tataris, and Tristan van Leeuwen for the valuable discussions that significantly contributed to the development of this study. Furthermore, we are grateful to the sponsors of the Utrecht Consortium for Subsurface Imaging (UCSI) for their financial funding and support.

\section*{Appendix A: Rank reduction in circulant matrices}\label{append_1}
Using a block diagonal cyclic-shear transformation, we derive the relationship between a circulant and a rank-one matrix. Consider the circulant matrix $\mathbf{C} \in \mathbb{R}^{n \times n}$ with associated polynomial
\begin{align*}
    \mathbf{C} = c_{1} \mathbf{\Pi}^{0} + c_{2} \mathbf{\Pi}^{1} + \cdots + c_{n} \mathbf{\Pi}^{n-1} \: ,
\end{align*}

where $\{ \mathbf{\Pi}^{i} \in \mathbb{R}^{n \times n} : 0 < i < n-1 \}$ is a set of cyclic shifts (\textit{Definition} \ref{def_2}), $\{ c_{i} \in \mathbb{R} : 1 < i < n \}$ is a generic coefficients set, and $\mathbf{\Pi}^{0} = \mathbf{I}_{n}$ is the identity matrix. An explicit series expansion results in a matrix of the form 
\begin{align*}
    \mathbf{C} = 
    \begin{bmatrix}
    c_{1}   & c_{n}   & \cdots & c_{3}  & c_{2}   \\
    c_{2}   & c_{1}   & c_{n}  &        & c_{3}   \\
    \vdots  & c_{2}   & c_{1}  & \ddots & \vdots  \\
    c_{n-1} &         & \ddots & \ddots & c_{n}   \\
    c_{n}   & c_{n-1} & \cdots & c_{2}  & c_{1}   \\
    \end{bmatrix} \: .
\end{align*}

As noted, the first column is the set of coefficients followed by a circular shift of itself, with each subsequent column being the circular shift of the previous one; thus, a single column or row fully determines the circulant. By virtue of the dominant column and row space misalignment, $\mathbf{C}$ is a high-rank matrix.

We now rearrange the entries $C_{ij}$ by applying the cyclic shear operator $\mathbf{S}^{\ast} = \mbox{diag}({\mathbf{\Pi}^{0}}^{\ast}, {\mathbf{\Pi}^{1}}^{\ast}, ..., {\mathbf{\Pi}^{n-1}}^{\ast}) \in \mathbb{R}^{nn \times nn}$ to $\mathbf{C}$, resulting in the $n \times n$ matrix,
\begin{align*}
    \mathbf{S}^{\ast} \mathbf{C} = c_{1} \mathbf{S}^{\ast} \mathbf{\Pi}^{0} + c_{2} \mathbf{S}^{\ast} \mathbf{\Pi}^{1} + \cdots + c_{n} \mathbf{S}^{\ast} \mathbf{\Pi}^{n-1} \: ,
\end{align*}

where $\mathbf{S}^{\ast}$ vectorizes the circulant matrix and applies the cyclic-shear map before reshaping the vector back into its matrix form. Next, let us evaluate the first term in the series:
\begin{align*}
    c_{1} \mathbf{S}^{\ast} \mathbf{\Pi}^{0} &= c_{1} \mbox{diag}({\mathbf{\Pi}^{0}}^{\ast}, {\mathbf{\Pi}^{1}}^{\ast}, ..., {\mathbf{\Pi}^{n-1}}^{\ast}) [\mathbf{e}^{T}_{1}, \mathbf{e}^{T}_{2}, ..., \mathbf{e}^{T}_{n-1}, \mathbf{e}^{T}_{n}]^{T}  \\
     &= c_{1} [{\mathbf{\Pi}^{0}}^{\ast}\mathbf{e}_{1}, {\mathbf{\Pi}^{1}}^{\ast}\mathbf{e}_{2}, ..., {\mathbf{\Pi}^{n-2}}^{\ast}\mathbf{e}_{n-1}, {\mathbf{\Pi}^{n-1}}^{\ast}\mathbf{e}_{n}] \\
     &= c_{1} [\mathbf{e}_{1}, \mathbf{e}_{1}, ..., \mathbf{e}_{1}, \mathbf{e}_{1}] \\
     &= c_{1} \mathbf{e}_{1} \otimes \mathbf{1}  \: .
\end{align*}

Here, it is assumed that the vector stack of the standard basis $\{ \mathbf{e}_{i} \in \mathcal{R}^{n} \}$ is the vectorized form of $\mathbf{\Pi}^{0}$, i.e., $\mbox{vec}(\mathbf{\Pi}^{0}) = [\mathbf{e}^{T}_{1}, \mathbf{e}^{T}_{2}, ..., \mathbf{e}^{T}_{n-1}, \mathbf{e}^{T}_{n}]^{T}$, $\mathbf{1} = [1, 1, ..., 1]^{T} \in \mathbb{R}^{n}$ is the all-ones vector, and $\otimes$ denotes the outer product. Likewise, evaluating the second term reduces to:  
\begin{align*}
    c_{2} \mathbf{S}^{\ast} \mathbf{\Pi}^{1} &= c_{2} \mbox{diag}({\mathbf{\Pi}^{0}}^{\ast}, {\mathbf{\Pi}^{1}}^{\ast}, ..., {\mathbf{\Pi}^{n-1}}^{\ast}) [\mathbf{e}^{T}_{2}, \mathbf{e}^{T}_{3}, ..., \mathbf{e}^{T}_{n}, \mathbf{e}^{T}_{1}]^{T}  \\
     &= c_{2} [{\mathbf{\Pi}^{0}}^{\ast}\mathbf{e}_{2}, {\mathbf{\Pi}^{1}}^{\ast}\mathbf{e}_{3}, ..., {\mathbf{\Pi}^{n-2}}^{\ast}\mathbf{e}_{n}, {\mathbf{\Pi}^{n-1}}^{\ast}\mathbf{e}_{1}] \\
     &= c_{2} [\mathbf{e}_{2}, \mathbf{e}_{2}, ..., \mathbf{e}_{2}, \mathbf{e}_{2}] \\
     &= c_{2} \mathbf{e}_{2} \otimes \mathbf{1} \: .
\end{align*}

Similarly, a consecutive examination of the remaining terms yields analogous results, with the last one reducing to:
\begin{align*}
    c_{n} \mathbf{S}^{\ast} \mathbf{\Pi}^{n-1} &= c_{n} \mbox{diag}({\mathbf{\Pi}^{0}}^{\ast}, {\mathbf{\Pi}^{1}}^{\ast}, ..., {\mathbf{\Pi}^{n-1}}^{\ast}) [\mathbf{e}^{T}_{n}, \mathbf{e}^{T}_{1}, ..., \mathbf{e}^{T}_{n-1}]^{T}  \\
     &= c_{n} [{\mathbf{\Pi}^{0}}^{\ast}\mathbf{e}_{n}, {\mathbf{\Pi}^{1}}^{\ast}\mathbf{e}_{1}, ..., {\mathbf{\Pi}^{n-2}}^{\ast}\mathbf{e}_{n-2}, {\mathbf{\Pi}^{n-1}}^{\ast}\mathbf{e}_{n-1}] \\
     &= c_{n} [\mathbf{e}_{n}, \mathbf{e}_{n}, ..., \mathbf{e}_{n}, \mathbf{e}_{n}] \\
     &= c_{n} \mathbf{e}_{n} \otimes \mathbf{1}  \: .
\end{align*}

Finally, grouping all the terms in the given linear combination leads to 
\begin{align*}
    \mathbf{S}^{\ast} \mathbf{C} &= c_{1} \mathbf{e}_{1} \otimes \mathbf{1} + c_{2} \mathbf{e}_{2} \otimes \mathbf{1} + ... + c_{n} \mathbf{e}_{n} \otimes \mathbf{1}  \\
     &= [c_{1}, c_{2}, ..., c_{n}] \otimes \mathbf{1} \\
     &= \mathbf{c}_{n} \otimes \mathbf{1} \: ,
\end{align*}

where $\mathbf{c}_{n} = [c_{1}, c_{2} ..., c_{n}]^{T} \in \mathbb{R}^{n}$ is the coefficients vector. In this case, the linear combination of rank-$1$ matrices of the form $c_{i} \mathbf{e}_{i} \otimes \mathbf{1}$, $i = 1, 2, ..., n$, is a rank-$1$ matrix $\mathbf{c}_{n} \otimes \mathbf{1}$. Therefore, $\mathbf{S}^{\ast}$ transforms the high-rank matrix $\mathbf{C}$ into a low-rank domain, where the data are rearranged into the rank-$1$ matrix, i.e., $\mathbf{S}^{\ast} \mathbf{C} = \mathbf{c}_{n} \otimes \mathbf{1}$. The rank-$1$ matrix is of the form 
\begin{align*}
    \mathbf{c}_{n} \otimes \mathbf{1} = 
    \begin{bmatrix}
    c_{1}  & c_{1} & c_{1}  & \cdots & c_{1} \\
    c_{2}  & c_{2} & c_{2}  &        & c_{2} \\
    c_{3}  & c_{3} & c_{3}  & \cdots & c_{3} \\
    \vdots &       & \vdots &        & \vdots  \\
    c_{n}  & c_{n} & c_{n}  & \cdots & c_{n}   \\
    \end{bmatrix} . 
\end{align*}

It is evident that such representation preserves the information in the input matrix while reducing its rank by aligning the rows and columns accordingly, ultimately revealing the low-dimension structure in the circulant matrix. Therefore, the cyclic shear serves as an alternative option to define a circulant matrix, 
\begin{align*}
    \mathbf{C} = \mathbf{S} \mathbf{c}_{n} \otimes \mathbf{1} \: . 
\end{align*}

Note that one is free to define the shift direction induced by any $\mathbf{\Pi}^{k}$ in $\mathbf{S}^{\ast}$ as long as the transformation matrix remains orthogonal.  

\bibliographystyle{unsrt}  
\bibliography{references}  






\end{document}